\documentclass[preprint,12pt]{elsarticle}
\usepackage[a4paper,left=2.6cm,right=2.6cm,top=2.5cm,bottom=2.5cm]{geometry}
\usepackage{amsmath,amssymb,amsfonts}
\usepackage{bm}
\usepackage{graphicx}
\usepackage{booktabs}
\usepackage{multirow}
\usepackage{array}
\usepackage{tabularx}
\usepackage{longtable}
\usepackage{makecell}
\usepackage{siunitx}
\usepackage{xcolor}
\usepackage{hyperref}
\usepackage{cleveref}
\usepackage{lineno}
\usepackage{float}
\usepackage{subcaption}
\usepackage{xurl}
\makeatletter
\g@addto@macro\UrlBreaks{\do\:\do\_}
\makeatother

\DeclareRobustCommand{\pyparam}[1]{\nolinkurl{#1}}

\DeclareRobustCommand{\pT}{\ensuremath{p_{\mathrm{T}}}}

\DeclareRobustCommand{\sqrts}{\ensuremath{\sqrt{s}}}

\DeclareRobustCommand{\mc}{\ensuremath{\mathrm{MC}}}
\DeclareRobustCommand{\data}{\ensuremath{\mathrm{data}}}

\DeclareRobustCommand{\ptzero}{\ensuremath{p_{\mathrm{T0}}}}

\DeclareRobustCommand{\ecmpow}{\pyparam{MultipartonInteractions:ecmPow}}
\DeclareRobustCommand{\ptref}{\pyparam{MultipartonInteractions:pT0Ref}}
\DeclareRobustCommand{\crrange}{\pyparam{ColourReconnection:range}}
\DeclareRobustCommand{\alund}{\pyparam{StringZ:aLund}}
\DeclareRobustCommand{\blund}{\pyparam{StringZ:bLund}}
\DeclareRobustCommand{\sigmaPT}{\pyparam{StringPT:sigma}}
\DeclareRobustCommand{\probStoUD}{\pyparam{StringFlav:probStoUD}}
\DeclareRobustCommand{\expPow}{\pyparam{MultipartonInteractions:expPow}}
\DeclareRobustCommand{\probQQtoQ}{\pyparam{StringFlav:probQQtoQ}}

\begin{document}

\begin{frontmatter}
\title{Sequential retuning of PYTHIA~8.316 to a global soft-QCD basis in pp collisions at $\sqrts=0.9$--$13$~TeV}

\author[1]{Haifa I. Alrebdi\corref{cor1}}
\ead{hialrebdi@pnu.edu.sa}
\author[2]{Muhammad Ajaz\corref{cor2}}
\ead{ajaz@awkum.edu.pk}

\cortext[cor2]{Corresponding author}
\address[1]{Department of Physics, College of Science, Princess Nourah bint Abdulrahman University, P.O. Box 84428, Riyadh 11671, Saudi Arabia}
\address[2]{Department of Physics, Abdul Wali Khan University Mardan, 23200 Mardan, Pakistan}

\begin{abstract}
We present a sequential nine-parameter retune of \textsc{PYTHIA}~8.316 constructed around the Monash~2013 baseline for soft-QCD observables in proton--proton collisions at $\sqrts=0.9$--$13$~TeV. The tune is built with direct generator evaluation rather than an interpolation surrogate, and its construction follows one continuous path: an initial five-parameter localization in the hadronization-sensitive sector, successive extensions driven by multi-energy minimum-bias tensions, and a final broadening to a global soft-QCD fit basis. The fitted observable set combines minimum-bias charged-particle measurements, identified-hadron spectra and ratios, underlying-event observables, and event-shape distributions. A separate held-out validation basis is also defined and kept outside the fit objective.

The final point retunes nine non-perturbative parameters controlling Lund fragmentation, strangeness suppression, diquark production, color reconnection, the MPI regularization scale, its energy evolution, the impact-parameter overlap profile, and transverse momentum generation in string breaking. On the common fit basis, the grouped score density improves from $S_{\mathrm{grouped}}/N_{\mathrm{bin}}=99.18$ for Monash~2013 to $85.67$ for the present tune. The strongest gains occur in the 13~TeV minimum-bias and underlying-event sectors, with further support from the ALICE 7 and 13~TeV identified-hadron blocks and a smaller but still positive improvement in the CMS 13~TeV identified-hadron sector. Monash remains better in the CMS 7~TeV underlying-event block and in the ATLAS 7~TeV event-shape block, while the largest unresolved tension for both tunes remains the low-energy charged-particle sector.

The held-out validation basis supports the same picture. Although Monash remains slightly better in the low-energy CMS validation blocks, the present tune performs clearly better in the ATLAS early underlying-event block and in both held-out 13~TeV charged-particle sectors. We therefore interpret the present tune not as a universal replacement for Monash, but as a controlled soft-QCD retune on the present global basis that improves a broad set of related observables while keeping its remaining limitations explicit.
\end{abstract}


\end{frontmatter}
%###################################

\section{Introduction}
\label{sec:introduction}

General-purpose Monte Carlo event generators are indispensable in collider phenomenology. They provide the only practical framework in which perturbative hard scattering, parton showers, multiparton interactions, beam remnants, hadronization, and unstable-particle decays are assembled into exclusive hadronic final states that can be compared directly with data \cite{Sjostrand:2008,Sjostrand:2015,Bierlich:2022Pythia83}. In proton--proton collisions this role is especially important in the soft and semisoft domain, where the underlying dynamics involve both perturbatively ordered radiation and genuinely non-perturbative physics. In that region, a prediction is inseparable from the parameter choices used to define the model \cite{Andersson:1983,Sjostrand:1987MPI,Sjostrand:2005Interleaved,Corke:2011,Buckley:2010Professor}. Event-generator tuning is therefore not just a technical adjustment. It is part of the physics content of the generator itself.

Within this class of tools, \textsc{PYTHIA} remains one of the standard references for soft-QCD studies at hadron colliders \cite{Sjostrand:2008,Sjostrand:2015,Bierlich:2022Pythia83}. Its hadronic final states are built from the Lund string picture of fragmentation \cite{Andersson:1983}, combined with an impact-parameter-based model for multiparton interactions and with color-reconnection mechanisms that alter the prehadronic string topology before fragmentation \cite{Sjostrand:1987MPI,Sjostrand:2005Interleaved,Corke:2011,Bierlich:2015CR,Christiansen:2015LC}. The parameters controlling these sectors influence charged-particle multiplicities, pseudorapidity densities, underlying-event observables, event-shape variables, flavor composition, and identified-hadron transverse-momentum spectra. None of these sectors can presently be fixed from first principles with the accuracy required for collider applications, so their description must be anchored to data.

The modern tuning program has benefited greatly from the availability of common comparison tools and public data repositories. \textsc{Rivet} provides a stable analysis layer for generator--data comparisons, while HEPData provides the corresponding numerical measurements in reusable form \cite{Bierlich:2020Rivet3,Bierlich:2024Rivet4,Maguire:2017HEPData}. These tools make it possible to define a tune basis clearly, preserve the exact comparison chain, and separate the physics logic of a tune from accidental details of implementation. They do not remove the need for physics judgment, but they do make the structure of a tuning study far more transparent than in earlier tune generations.

Among \textsc{PYTHIA}~8 tunes, Monash~2013 remains the natural baseline for any non-perturbative retuning exercise \cite{Skands:2014Monash}. Its construction combined revised fragmentation constraints with selected hadron-collision measurements and produced a well-balanced tune that improved several sectors simultaneously \cite{Skands:2014Monash}. At the same time, Monash was never intended to close the tuning problem. Its own discussion pointed to remaining tensions, especially in strange-particle production, baryon-sensitive observables, and some low-\(\pT\) structures \cite{Skands:2014Monash}. Since then, the LHC soft-QCD dataset has expanded substantially. Minimum-bias charged-particle measurements now span a broader energy range, identified-hadron spectra are available at both 7 and 13~TeV with improved coverage, and dedicated underlying-event and event-shape measurements offer direct constraints on event activity and topology \cite{Aamodt:2010nch900236,Aamodt:2010nch7,Adam:2015pp7pid,Khachatryan:2015eta13,Sirunyan:2017pp13pid,Acharya:2021LightFlavor7and13,Aad:2011UE,Khachatryan:2011UE,Aad:2013EventShapes,Aaboud:2017UE13,Aad:2016MB13,Aad:2016TrackMB13,CMS:2018ChargedFS13}. A tune that remains adequate in one subset of these observables does not automatically remain adequate in the others. This broader motivation is consistent with recent generator-comparison studies by the present collaboration on charged-particle spectra and pseudorapidity densities, track-based charged-particle observables, multiplicity-dependent jet and underlying-event observables, 13~TeV underlying-event-sensitive charged-particle distributions, and 0.9~TeV charged-particle spectra, all of which showed marked sector-to-sector differences in model performance in pp collisions \cite{Ajaz:2023EntropyCharged,Ajaz:2023SymmetryTrack,Waqar:2024CPCUE,Alrebdi:2025EPJPUE13,Alrebdi:2024CJP09TeV}.

This is the main motivation for the present work. The aim is not to produce another one-shot high-dimensional scan in which the origin of every gain becomes opaque. Instead, we construct the final tune sequentially. The active parameter set is enlarged only when the residual structure of the score shows that the current subspace is too restrictive. The observable basis is also broadened in steps, moving from a compact hadronization-sensitive core toward a wider soft-QCD basis that includes minimum-bias, identified-hadron, underlying-event, and event-shape information. That strategy keeps the tune readable. It also makes it possible to see which newly opened degree of freedom responds to which sector of the data. In that sense, the present workflow is complementary to established tune-construction strategies based on expert tune families, interpolation-based tuning, Bayesian optimization, and more recent high-dimensional or color-reconnection-focused studies \cite{Skands:2010Perugia,Buckley:2010Professor,Ilten:2017BO,Krishnamoorthy:2021Apprentice,Bellm:2020HighDim,Tumasyan:2023CRtunes}.

The result of that construction is a nine-parameter retune of \textsc{PYTHIA}~8.316 built around the Monash 2013 baseline. The final fit basis spans pp collisions at
\[
\sqrt{s}=0.9,\ 2.36,\ 7,\ \text{and }13~\mathrm{TeV},
\]
and includes low-energy and intermediate-energy charged-particle measurements, identified-hadron spectra and ratios, 7~TeV and 13~TeV underlying-event observables, 7~TeV event-shape variables, and 13~TeV track-based minimum-bias distributions. In parallel, a held-out validation basis is defined and kept outside the fit objective. This distinction is important. The present paper does not report only a fit-basis improvement relative to Monash. It also checks whether the tuned point remains competitive on observables that were not used to determine it.

The physics goal of this broader basis is straightforward. Minimum-bias charged-particle observables constrain the normalization and energy evolution of soft activity and are therefore sensitive to the multiparton-interaction sector and its overlap profile \cite{Sjostrand:1987MPI,Sjostrand:2005Interleaved,Corke:2011,Aamodt:2010nch900236,Khachatryan:2010ChargedPtEta,Aad:2016MB13,Aad:2016TrackMB13}. Identified-hadron spectra and ratios probe fragmentation hardness, flavor composition, strangeness suppression, and baryon sensitivity in a much more differential way \cite{Adam:2015pp7pid,Sirunyan:2017pp13pid,Acharya:2021LightFlavor7and13}. Underlying-event observables constrain how additional activity populates the transverse and toward/away regions relative to a leading object \cite{Aad:2011UE,Khachatryan:2011UE,Aaboud:2017UE13}. Event-shape variables test whether the same tune also describes the broader topology of soft hadronic final states \cite{Aad:2013EventShapes}. A tune that improves several of these sectors simultaneously is more informative than one that only optimizes a single observable family. For convenience, we refer to the final parameter set obtained in this study as the \emph{Ajaz tune}.

The present paper is therefore organized around one question: how far can a controlled sequential retuning of a compact but physically motivated \textsc{PYTHIA} subspace improve the global soft-QCD description relative to Monash 2013? Section~\ref{sec:data} defines the fit basis and the held-out validation basis. Section~\ref{sec:setup} summarizes the generator setup and the nine active parameters of the final tune. Section~\ref{sec:strategy} describes the direct-generation workflow, the grouped score, and the stepwise construction that leads from the first five-parameter localization to the final global Ajaz tune. Section~\ref{sec:results} presents the final benchmark against Monash on the common fit basis, including the held-out validation comparison. Section~\ref{sec:discussion} examines the gains and residual tensions on an analysis-by-analysis basis, while Section~\ref{sec:conclusion} collects the main conclusions.

\section{Experimental input and analysis basis}
\label{sec:data}

\subsection{Fit basis}
\label{subsec:fit_basis}

The final fit basis was constructed to constrain the principal soft-QCD structures addressed by the active parameter set, while avoiding an indiscriminate all-observable tune. The selected analyses cover four center-of-mass energies and four closely connected physics sectors: minimum-bias charged-particle production, identified-hadron production, underlying-event activity, and event-shape observables. 

For readability, the main text refers to these analyses by short physics-style labels rather than by repeating the raw Rivet identifiers. The mapping used throughout the paper is given in Table~\ref{tab:analysis_map}. The full fit basis is listed in Table~\ref{tab:fit_basis}.
 
The low and intermediate-energy charged-particle baseline is given by the ALICE charged-particle block, which provides charged-particle multiplicity and pseudorapidity information at $\sqrt{s}$ = 0.9, 2.36, and 7~TeV \cite{Aamodt:2010nch900236,Aamodt:2010nch7}. This block plays two roles. First, it constrains the overall normalization and energy transport of the minimum-bias sector across a wide energy range. Second, it provides a direct test of whether a tune that works well in the 7 and 13~TeV sectors remains stable when transported downward to the lower-energy region.

The 7~TeV identified-hadron sector is represented by the ALICE 7 TeV identified-hadron block, which provides \(\pi^\pm\), \(K^\pm\), \(p\), and \(\bar p\) spectra together with flavor-sensitive ratios \cite{Adam:2015pp7pid}. This block is central to the early tuning stages because it responds directly to the Lund fragmentation parameters, strangeness suppression, color reconnection, and later to the transverse-smearing freedom. It is also the cleanest place in the fit basis to see whether the same point reproduces both spectral shape and relative flavor composition. The 7~TeV soft-activity and topology sectors are extended with two additional analyses. The first is the CMS 7 TeV UE block, which constrains charged-particle and scalar-\(\pT\) densities in the standard toward, away, and transverse regions relative to the leading jet \cite{Khachatryan:2011UE}. The second is the ATLAS 7 TeV event-shape block, which provides transverse thrust, thrust minor, and transverse sphericity distributions in minimum-bias events \cite{Aad:2013EventShapes}. These observables were introduced only after the tune had already stabilized on the narrower mixed-family basis. Their role in the final fit is to ensure that the tuned point is not only locally successful in spectra and densities, but also reasonably consistent with global event geometry.

The 13~TeV sector is intentionally broader than the earlier mixed-family draft. It begins with the CMS 13 TeV charged-density block \cite{Khachatryan:2015eta13}, which constrains the inclusive charged-particle normalization at the highest energy used in the fit. It is then enlarged by two identified-hadron blocks: the CMS 13 TeV identified-hadron block, which measures charged pions, kaons, and protons \cite{Sirunyan:2017pp13pid}, and the ALICE 13 TeV light-flavor block, which broadens the 13~TeV flavor information beyond the CMS selection and adds an independent light-flavor reference at the same energy \cite{Acharya:2021LightFlavor7and13}. These two analyses are especially important for separating purely pionic improvements from genuine flavor-composition improvements in the kaon and baryon sectors.

\begin{table}[H]
\centering
\caption{Mapping between the short analysis labels used in the text and the corresponding Rivet analyses.}
\label{tab:analysis_map}
\begin{tabularx}{\textwidth}{>{\raggedright\arraybackslash}p{0.23\textwidth} X >{\centering\arraybackslash}p{0.10\textwidth} >{\raggedright\arraybackslash}p{0.22\textwidth}}
\toprule
Short label & Measurement & $\sqrts$ (TeV) & Rivet analysis \\
\midrule
ALICE charged-particle block & ALICE minimum-bias charged-particle densities and multiplicities & 0.9, 2.36, 7 & \texttt{ALICE\_2010\_I852264} \\
ALICE 7 TeV identified-hadron block & ALICE \(\pi^\pm\), \(K^\pm\), \(p\), and \(\bar p\) spectra and flavor ratios & 7 & \texttt{ALICE\_2015\_I1357424} \\
CMS 7 TeV UE block & CMS underlying-event observables & 7 & \texttt{CMS\_2011\_I916908} \\
ATLAS 7 TeV event-shape block & ATLAS charged-particle event-shape observables & 7 & \texttt{ATLAS\_2012\_I1124167} \\
CMS 13 TeV charged-density block & CMS charged-hadron pseudorapidity density & 13 & \texttt{CMS\_2015\_I1384119} \\
CMS 13 TeV identified-hadron block & CMS \(\pi^\pm\), \(K^\pm\), \(p\), and \(\bar p\) spectra and ratios & 13 & \texttt{CMS\_2017\_I1608166} \\
ALICE 13 TeV light-flavor block & ALICE light-flavor hadron production & 13 & \texttt{ALICE\_2020\_I1797443} \\
ATLAS 13 TeV MB track block & ATLAS low-\(\pT\) minimum-bias charged-particle distributions & 13 & \texttt{ATLAS\_2016\_I1467230} \\
ATLAS 13 TeV UE block & ATLAS track-based underlying-event observables & 13 & \texttt{ATLAS\_2017\_I1509919} \\
CMS low-energy charged-particle block & CMS charged-particle \(\pT\) and \(\eta\) distributions & 0.9, 2.36, 7 & \texttt{CMS\_2010\_I845323} \\
CMS low-energy NSD block & CMS non-single-diffractive charged-particle multiplicities & 0.9, 2.36, 7 & \texttt{CMS\_2011\_I879315} \\
ATLAS early UE block & ATLAS underlying-event observables & 0.9, 7 & \texttt{ATLAS\_2010\_I879407} \\
ATLAS 13 TeV charged-track validation block & ATLAS charged-particle distributions & 13 & \texttt{ATLAS\_2016\_I1419652} \\
CMS 13 TeV charged-final-state block & CMS charged-particle distributions in different final states & 13 & \texttt{CMS\_2018\_I1680318} \\
\bottomrule
\end{tabularx}
\end{table}

The final two fit-basis entries strengthen the 13~TeV activity sector beyond the single CMS pseudorapidity-density observable. The ATLAS 13 TeV MB track block provides track-based minimum-bias distributions with low-\(\pT\) charged particles \cite{Aad:2016TrackMB13}, while the ATLAS 13 TeV UE block provides a track-based underlying-event measurement at the same energy \cite{Aaboud:2017UE13}. Their inclusion is one of the defining differences between the final Ajaz tune and the earlier seven-parameter mixed-family benchmark. These analyses make the final fit basis genuinely multi-sector at 13~TeV rather than only multi-observable in a narrow sense.

It is not a universal basis, but it covers the main soft-QCD structures that are directly affected by the active parameter set.

\begin{table}[H]
\centering
\caption{Fit-basis analyses used to determine the final Ajaz tune. Exact Rivet identifiers are mapped in Table~\ref{tab:analysis_map}.}
\label{tab:fit_basis}
\begin{tabularx}{\textwidth}{>{\raggedright\arraybackslash}p{0.28\textwidth} >{\centering\arraybackslash}p{0.12\textwidth} >{\raggedright\arraybackslash}p{0.20\textwidth} X}
\toprule
Short label & $\sqrts$ (TeV) & Sector & Main role in the tune \\
\midrule
ALICE charged-particle block & 0.9, 2.36, 7 & minimum-bias charged particles & low-energy and 7 TeV charged-particle anchor \\
ALICE 7 TeV identified-hadron block & 7 & identified hadrons & pions, kaons, protons, flavor-sensitive ratios \\
CMS 7 TeV UE block & 7 & underlying event & UE activity and transverse-region structure \\
ATLAS 7 TeV event-shape block & 7 & event shapes & global event topology at 7 TeV \\
CMS 13 TeV charged-density block & 13 & charged-particle density & high-energy charged-particle normalization and shape \\
CMS 13 TeV identified-hadron block & 13 & identified hadrons & 13 TeV light-hadron spectra and ratios \\
ALICE 13 TeV light-flavor block & 13 & light-flavor hadrons & 13 TeV flavor extension beyond the CMS block \\
ATLAS 13 TeV MB track block & 13 & minimum-bias tracks & multiplicity-sensitive track observables \\
ATLAS 13 TeV UE block & 13 & underlying event & track-based UE constraints at 13 TeV \\
\bottomrule
\end{tabularx}
\end{table}

\subsection{Held-out validation basis}
\label{subsec:validation_basis}

A second analysis block was reserved for validation and was excluded from the fit objective throughout the final tuning campaign. The purpose of this held-out basis is not to create a mathematically orthogonal observable set. That would be unrealistic in soft-QCD tuning, where neighboring minimum-bias and underlying-event measurements often probe closely related physics. The purpose is instead to keep a meaningful set of independent analyses outside the score used to select the final point, so that the tuned result can be tested on nearby but distinct measurements.

The validation basis is listed in Table~\ref{tab:validation_basis}. It contains two CMS charged-particle analyses at 0.9, 2.36, and 7~TeV, namely the CMS low-energy charged-particle block \cite{Khachatryan:2010ChargedPtEta} and the CMS low-energy NSD block \cite{Khachatryan:2011NSD}, together with the ATLAS early UE block at 0.9 and 7~TeV \cite{Aad:2011UE}. The 13~TeV validation sector contains the ATLAS 13 TeV charged-track validation block \cite{Aad:2016MB13} and the CMS 13 TeV charged-final-state block \cite{CMS:2018ChargedFS13}. None of these analyses contributed to the fit-basis score that selected the final Ajaz point.

This split between fit and validation serves two purposes. First, it tests whether the tuned point generalizes beyond the exact histograms that determined it. Second, it avoids the common weakness of tune studies that report only improvements on the same basis used during optimization. The analysis is therefore organized around two comparisons: a fit-basis benchmark against Monash and a held-out validation benchmark against the same reference.

\begin{table}[H]
\centering
\caption{Held-out validation analyses. These observables were excluded from the fit objective. Exact Rivet identifiers are mapped in Table~\ref{tab:analysis_map}.}
\label{tab:validation_basis}
\begin{tabularx}{\textwidth}{>{\raggedright\arraybackslash}p{0.28\textwidth} >{\centering\arraybackslash}p{0.12\textwidth} >{\raggedright\arraybackslash}p{0.20\textwidth} X}
\toprule
Short label & $\sqrts$ (TeV) & Sector & Validation role \\
\midrule
CMS low-energy charged-particle block & 0.9, 2.36, 7 & minimum-bias charged particles & independent low-energy charged-particle check \\
CMS low-energy NSD block & 0.9, 2.36, 7 & NSD charged particles & validation of neighboring MB selection \\
ATLAS early UE block & 0.9, 7 & underlying event & early UE cross-check outside the fit basis \\
ATLAS 13 TeV charged-track validation block & 13 & charged-particle distributions & independent 13 TeV track-based test \\
CMS 13 TeV charged-final-state block & 13 & charged particles in different final states & high-energy validation of track activity and event selection dependence \\
\bottomrule
\end{tabularx}
\end{table}

\subsection{Event-class consistency and sector organization}
\label{subsec:event_classes}

A combined soft-QCD fit basis only makes sense if incompatible event definitions are not mixed carelessly inside one common score. That issue is especially important in the minimum-bias sector, where inelastic and non-single-diffractive selections are close in subject matter but not identical in physics content. It is also important in the underlying-event sector, where the presence of a leading charged particle or leading jet defines a different event environment from the one probed by inclusive minimum-bias measurements.

The present tune handles this by grouping observables analysis by analysis. The primary score is built from analysis-block contributions rather than by merging all histogram bins into an undifferentiated pool. This preserves the experimental selection boundaries and makes it possible to identify which sector is responsible for a gain or a deficit. It also avoids interpreting a local improvement in one event class as a general improvement of a neighboring class.

The fit basis was assembled accordingly. The inelastic ALICE charged-particle block was retained in the fit because it constrains the energy evolution of inclusive charged production. The CMS low-energy NSD block, while clearly relevant, was placed in the validation basis instead of being folded directly into the fit objective. The logic is simple: it remains scientifically valuable, but it tests the tuned point more cleanly when left outside the minimization target.

The same care was taken in the later broadening of the fit basis. Underlying-event and event-shape analyses were not inserted into the earliest tune stages, because doing so too soon would have obscured the interpretation of the first hadronization-sensitive basin. They were added only after the tune had already stabilized on the narrower mixed-family basis. In that sense, the fit basis is broad, but not indiscriminate. Each addition was made at a point where the existing basin could be tested meaningfully against the new sector.

The sector organization adopted in Table~\ref{tab:fit_basis} therefore reflects a methodological decision as well as a physics one. The final Ajaz tune is constrained by several classes of observables, but each class remains identifiable inside the grouped score. That is essential for the interpretation of the results in Sections~\ref{sec:results} and \ref{sec:discussion}.

% =========================================================
\section{Generator setup and active tune parameters}
\label{sec:setup}

\subsection{Baseline generator configuration}
\label{subsec:baseline_setup}

All events in the present study were generated with \textsc{PYTHIA}~8.316 and analyzed with \textsc{Rivet}~4.1.1. The baseline configuration was the Monash~2013 tune \cite{Skands:2014Monash}. This is the correct reference point for the present work because the final Ajaz tune remains inside the same broad non-perturbative modeling framework: Lund string fragmentation, color reconnection, and multiparton interactions are all retained, and the retune acts only on a restricted subset of parameters inside that framework \cite{Bierlich:2022Pythia83,Skands:2014Monash}.

The final Ajaz tune is therefore not a new stand-alone generator setup. It is an incremental retune relative to Monash. All settings not explicitly promoted into the active parameter set were kept at their Monash values. This keeps the comparison interpretable and avoids a common ambiguity in tune studies, namely that several settings are changed at once relative to the reference, making it unclear whether an observed improvement can be traced to a specific retuning step or to a broader uncontrolled change in the generator configuration.

The workflow was based on direct event generation at explicit parameter points rather than on a response-surface interpolation. For each tune point, we prepared a dedicated \textsc{PYTHIA} command card, generated statistically independent replicas at the relevant collision energies, and passed the merged outputs through the same \textsc{Rivet} analysis chain used for Monash. All comparisons in the paper are therefore made on the same footing: the generator version, analysis implementation, stage-level event statistics, and score definition are all kept fixed.

The final tune is constrained across
\[
\sqrt{s}=0.9,\ 2.36,\ 7,\ \text{and }13~\mathrm{TeV}.
\]
These energies play distinct roles in the fit. The 0.9 and 2.36~TeV points probe the low-energy transport of the soft-QCD model. The 7~TeV point serves as the main bridge, where minimum-bias, identified-hadron, underlying-event, and event-shape information all enter simultaneously. The 13~TeV point tests the same parameter set at the highest energy included in the tune and is where the broader soft-activity and flavor-sensitive sectors were added most strongly.

\subsection{Nine active parameters and their physics roles}
\label{subsec:active_parameters}

The final Ajaz tune is defined in a nine-parameter non-perturbative subspace. These parameters were not activated at once. They were promoted into the active set as the fit basis widened and as the residual score pattern showed which additional degree of freedom was needed. Their physics roles are summarized in Table~\ref{tab:active_parameters}.

The pair \alund\ and \blund\ controls the longitudinal shape of the Lund symmetric fragmentation function. These are the primary shape parameters of string fragmentation and influence the hardness of the identified-hadron spectra and their species-dependent response \cite{Bierlich:2022Pythia83,Andersson:1983,Skands:2014Monash}. They were part of the active set from the very first localization stage because no identified-hadron tune can be defined sensibly while keeping both fixed.

The parameter \probStoUD\ controls the suppression of strange-quark pair production during string breaking. It is directly exposed by kaon production and by flavor-sensitive hadron ratios \cite{Andersson:1983,Skands:2014Monash}. The later-added parameter \probQQtoQ\ controls diquark-pair production relative to quark production and is therefore one of the main handles on baryon-sensitive structure in the string model \cite{Andersson:1983,Skands:2014Monash,Bierlich:2022Pythia83}. It was kept fixed in the early stages and activated only when the fit basis had expanded enough for baryon-sensitive and heavier-species tensions to become a real limitation.

The parameter \crrange\ controls the strength of color reconnection. In practice, it influences the topology and effective length of strings before hadronization and therefore affects both multiplicity-related observables and identified-hadron production \cite{Skands:2014Monash,Christiansen:2015LC,Bierlich:2015CR}. This makes it one of the natural bridge parameters between the minimum-bias and identified-hadron sectors.

Three active parameters belong to the multiparton-interaction sector. The first, \ptref, is the reference infrared regularization scale that controls the onset and overall level of additional soft partonic activity \cite{Sjostrand:1987MPI,Sjostrand:2005Interleaved,Corke:2011}. The second, \ecmpow, controls the energy scaling of that regularization scale and becomes essential as soon as the tune basis becomes genuinely multi-energy \cite{Skands:2014Monash,Sirunyan:2020CP5}. The third, \expPow, shapes the overlap profile relevant to MPI activity and therefore provides a more direct handle on unresolved minimum-bias and underlying-event structure once simple normalization and energy transport are no longer enough \cite{Bierlich:2022Pythia83,Sjostrand:1987MPI,Sjostrand:2005Interleaved}.

\begin{table}[H]
\centering
\caption{Active tune parameters used in the final Ajaz tune and their main physics roles.}
\label{tab:active_parameters}
\begin{tabularx}{\textwidth}{>{\raggedright\arraybackslash}p{0.38\textwidth} >{\centering\arraybackslash}p{0.18\textwidth} X}
\toprule
Parameter & Short name & Primary role \\
\midrule
\alund & $a_{\mathrm{Lund}}$ & longitudinal string-fragmentation shape \\
\blund & $b_{\mathrm{Lund}}$ & longitudinal string-fragmentation shape \\
\probStoUD & $P_{s/u,d}$ & strangeness suppression in string breaking \\
\probQQtoQ & $P_{qq/q}$ & diquark-pair production and baryon sensitivity \\
\crrange & CR range & color-reconnection strength and topology \\
\ptref & $\ptzero^{\mathrm{ref}}$ & MPI infrared regularization scale \\
\ecmpow & $e_{\mathrm{pow}}$ & energy scaling of the MPI regularization scale \\
\expPow & expPow & MPI overlap-profile shape \\
\sigmaPT & $\sigma_{\pT}$ & transverse momentum width in string breaking \\
\bottomrule
\end{tabularx}
\end{table}

The final active parameter, \sigmaPT, controls the transverse momentum width generated during string breaking. It provides limited additional freedom in the soft broadening of hadron momenta beyond what can be achieved through the longitudinal fragmentation parameters alone \cite{Andersson:1983,Skands:2014Monash,Bierlich:2022Pythia83}. In the present tune it remained a secondary hadronization lever rather than a primary driver of the fit.

\subsection{Parameter-promotion philosophy}
\label{subsec:promotion_logic}
A central feature of the Ajaz tune construction is that the active parameter set was not chosen once at the beginning and then scanned blindly. Parameters were promoted only when both the physics content of the fit basis and the residual score pattern justified doing so. This promotion rule is one of the main reasons the tune remains readable even after expanding to a nine-parameter space.

The first localized basin was built in a compact five-parameter core,

\{\alund,\ \blund,\ \crrange,\ \ptref,\ \probStoUD\},

which was sufficient to test whether the identified-hadron sector, together with the basic charged-particle constraints, could support a viable local minimum. Once the basis became genuinely multi-energy, that core was no longer flexible enough. The next parameter promoted was naturally \ecmpow, since the dominant new tension came from the transport of soft activity from 7~TeV to both lower and higher energies. After the six-parameter basin had stabilized, \sigmaPT\ was introduced as a limited extra hadronization freedom to test whether additional transverse broadening could improve the mixed-family description.

The later broadening of the fit basis shifted the balance again. Once tensions in the minimum-bias and underlying-event sectors remained visible after the first seven-parameter stabilization, \expPow\ was activated because it gives more direct control of the MPI overlap profile than \ptref\ and \ecmpow\ alone. Finally, after the 13~TeV flavor basis had been enlarged and the heavier-species sector still showed a clear weakness, \probQQtoQ\ was promoted to introduce an explicitly baryon-sensitive degree of freedom.

This promotion strategy serves two purposes. It suppresses unnecessary degeneracy in the early stages of the tune, and it ties each added parameter to a concrete sector-level tension. The final nine-parameter Ajaz tune should therefore be read as the endpoint of a controlled sequence of parameter promotions, not as the result of a single undifferentiated high-dimensional scan.

% =========================================================

\section{Sequential construction of the Ajaz tune}
\label{sec:strategy}

\subsection{Direct-generation workflow and goodness-of-fit}
\label{subsec:workflow_gof}

The Ajaz tune was obtained from direct event generation at explicit parameter points. No interpolation surrogate was used. For each candidate point, we prepared a dedicated \textsc{PYTHIA} command card, generated statistically independent event samples at the relevant collision energies, and analyzed the merged outputs with the same \textsc{Rivet} routines used for the Monash reference. This point-by-point approach is slower than interpolation-based tuning, but in the present study it has a practical advantage: every quoted gain comes from an explicitly generated parameter point rather than from a fitted response surface \cite{Buckley:2010Professor,Skands:2014Monash}.

The tuning loop itself stayed unchanged throughout the study. We defined a bounded parameter subspace, sampled a finite set of points inside it, generated each point with several independent replicas, compared the merged histograms to data with a common grouped score, and then restricted the next scan to the most promising local basin. This cycle was repeated until the ranking stabilized. The active parameter set and the fit basis were then enlarged only when the residual score pattern showed that the current setup had become too restrictive.

At every scan point, replica merging was used rather than a single event stream. This reduced the influence of accidental Monte Carlo fluctuations on the ranking and produced one merged analysis output per physical point. The early exploratory stages used modest statistics per point because their purpose was basin finding, not publication-level benchmarking. The first five-parameter pilot scan used 15 replicas of \(2\times 10^5\) events per point, corresponding to
\[
15 \times 2\times 10^5 = 3.0\times 10^6
\]
events per point. The first local refinement used 12 replicas of \(2\times 10^5\) events per point, corresponding to
\[
12 \times 2\times 10^5 = 2.4\times 10^6
\]
events per point. The retained 7~TeV comparison stage used 12 replicas of \(2\times 10^5\) events per point, that is
\[
12 \times 2\times 10^5 = 2.4\times 10^6
\]
events per point. The final report-level Ajaz--Monash comparison was carried out with 24 replicas of \(2\times 10^5\) events per point,
\[
24 \times 2\times 10^5 = 4.8\times 10^6,
\]
so the quoted benchmark was obtained with substantially higher statistics than the exploratory scans.

The score was built directly from the binned data--MC comparison returned by \textsc{Rivet}. For each populated bin \(i\),
\begin{equation}
\chi_i^2 =
\frac{\left(y_i^{\mc}-y_i^{\data}\right)^2}
{\left(\sigma_i^{\data}\right)^2+\left(\sigma_i^{\mc}\right)^2},
\label{eq:chi_bin}
\end{equation}
where \(y_i^{\mc}\) and \(y_i^{\data}\) are the Monte Carlo and experimental central values and \(\sigma_i^{\mc}\) and \(\sigma_i^{\data}\) are the corresponding uncertainties. For each analysis block \(a\), the bin contributions were summed,
\begin{equation}
S_a = \sum_{i \in a}\chi_i^2,
\label{eq:score_block}
\end{equation}
and the grouped score for a full tune point was then constructed as
\begin{equation}
S_{\mathrm{grouped}} = \sum_a w_a S_a.
\label{eq:score_grouped}
\end{equation}
In the present study all active fit-basis analyses were assigned unit weights, \(w_a=1\). The point ranking was quoted as the grouped score density, \(S_{\mathrm{grouped}}/N_{\mathrm{bin}}\), where \(N_{\mathrm{bin}}\) denotes the number of populated bins contributing to the grouped comparison. 

\begin{table}[H]
\centering
\caption{Summary of the sequential tune-construction steps. Event counts refer to merged statistics per point when explicitly fixed.}
\label{tab:strategy_summary}
\small
\begin{tabularx}{\textwidth}{>{\raggedright\arraybackslash}p{0.18\textwidth} >{\raggedright\arraybackslash}p{0.22\textwidth} >{\centering\arraybackslash}p{0.14\textwidth} >{\centering\arraybackslash}p{0.16\textwidth} X}
\toprule
Step & Active parameters & Points & Events per point & Main purpose \\
\midrule
Initial localization & 5 parameters & 21 & \(3.0\times10^6\) & Identify the first viable basin in the hadronization-sensitive sector \\
Local refinement & same 5 parameters & 10 & \(2.4\times10^6\) & Tighten the preferred local basin \\
Retained 7~TeV test & same 5 parameters & 6 finalists & \(2.4\times10^6\) & Check the basin against a broader 7~TeV observable mix \\
Multi-energy mixed-family extension & 5, then 6 with \ecmpow & 12, then 6 retained & stage-dependent & Test compatibility with the low- and high-energy charged-particle sectors \\
Limited fragmentation broadening & 7 parameters with \sigmaPT & 12, then local refinement & stage-dependent & Check whether extra soft broadening improves the mixed-family basin \\
MPI overlap-profile opening & 8 parameters with \expPow & stage-dependent & stage-dependent & Address persistent minimum-bias and underlying-event tension \\
13~TeV flavor broadening & 9 parameters with \probQQtoQ & stage-dependent & stage-dependent & Add explicit baryon-sensitive freedom and enlarge the hadron basis \\
Final global basis construction & same 9 parameters & stage-dependent & stage-dependent & Bring UE, event-shape, and 13~TeV track-based MB observables into the fit basis \\
Report-level rerun & frozen Ajaz point vs Monash & 2 comparison points & \(4.8\times10^6\) & Final like-for-like benchmark on the common global basis \\
\bottomrule
\end{tabularx}
\end{table}
At the analysis-block level, we quote the corresponding blockwise weighted score density
$S_a/N_{\mathrm{bin},a}$, where $S_a$ is the weighted score of analysis block $a$ and $N_{\mathrm{bin},a}$ is the number
of populated bins in that block.

This quantity should be read as a controlled ranking metric for the chosen basis, not as a formal covariance-based reduced \(\chi^2\). Bin-to-bin experimental correlations were not included in the present workflow, so the score is built from uncorrelated bin contributions. The benchmark remains meaningful because the same definition, statistics policy, generator version, and analysis chain are used for every candidate point and for Monash itself.

\subsection{Initial five-parameter localization}
\label{subsec:five_param_localization}
The first step of the tuning programme was deliberately narrow. The active parameter set was
\{\alund,\ \blund,\ \crrange,\ \ptref,\ \probStoUD\}
which spans the longitudinal fragmentation shape, the main strangeness-suppression lever, one color-reconnection control, and the reference scale of the MPI sector. This five-parameter core was enough for the first practical test of the study: whether a compact and physically sensible non-perturbative subspace already contained a viable basin for the identified-hadron sector and the most basic charged-particle constraints.

The initial exploratory scan used 21 points which include one central and 20 additional points in the allowed domain. It was used to find the best region of the five-parameter space for local refinement. Two nearby points gave the best local basin, and this basin defined the reduced search region for the next step.

\subsection{Local refinement and retained finalists}
\label{subsec:local_refinement}

The next step refined the strongest basin found in the exploratory scan. Rather than launching another broad search, the active box was contracted to a reduced hyper-rectangle defined by the envelope of the leading pilot candidates, with a modest padding kept inside the original allowed domain. The active refinement intervals were
\begin{equation}
\begin{aligned}
\text{\texttt{StringZ:aLund}} &\in [0.48878204,\;1.2], \\
\text{\texttt{StringZ:bLund}} &\in [0.97046456,\;1.8772126], \\
\text{\texttt{ColourReconnection:range}} &\in [2.9537838,\;3.2772971], \\
\text{\texttt{MultipartonInteractions:pT0Ref}} &\in [2.313851,\;2.3910248], \\
\text{\texttt{StringFlav:probStoUD}} &\in [0.19915096,\;0.24529151].
\end{aligned}
\label{eq:refinement_box}
\end{equation}

This refinement used 10 points in total: the anchor points that defined the local basin and a smaller set of new samples inside the restricted box. The aim here was not to find a qualitatively different solution. It was to determine whether the leading basin from the pilot scan remained preferred once the local region was resolved more finely. It did.

After this refinement, exact duplicates between the exploratory and refined point sets were collapsed and a set of six unique retained finalists was constructed. These retained points then served two purposes. First, they preserved the local structure of the best current basin. Second, they provided the anchors from which the broader multi-energy construction could proceed without discarding the information already learned from the hadronization-sensitive start.

\subsection{Mixed-family multi-energy extension}
\label{subsec:mixed_family_extension}

The next question was whether the locally preferred five-parameter basin survived once the observable basis was widened. The first widening remained conservative. The retained finalists were first tested on a 7~TeV basis that combined the ALICE 7 TeV identified-hadron block with the 7~TeV charged-particle component of the ALICE charged-particle block. This was the first point in the study at which flavor-sensitive spectra and multiplicity-sensitive charged-particle information were forced to coexist inside the same score.

The basis was then promoted to a genuinely multi-energy mixed-family objective by combining
the ALICE charged-particle block at 0.9, 2.36, and 7~TeV with
the ALICE 7 TeV identified-hadron block at 7~TeV and
the CMS 13 TeV charged-density and identified-hadron blocks.
A local overlap scan was built from four anchor points taken from the retained five-parameter basin together with eight new samples drawn inside the same local envelope. This was the first real compatibility test between the hadronization-sensitive basin and the multi-energy charged-particle sector.

That test failed in a useful way. The previously acceptable five-parameter basin did not transport well to the full multi-energy mixed-family objective. The dominant deterioration came from the low-energy and global charged-particle sector, especially the ALICE charged-particle block, rather than from the 13~TeV identified-hadron sector. That observation determined the next step. A further five-parameter refinement would have sharpened the wrong subspace. What was needed was not more local sampling in the same coordinates, but one additional parameter that could control the energy transport of the soft-activity sector.

\subsection{Activation of \ecmpow}
\label{subsec:ecmpow_activation}

The natural next promotion was \ecmpow. Once the tune basis became truly multi-energy, the dominant residual mismatch could no longer be understood as a purely local 7~TeV hadronization problem. In \textsc{PYTHIA}, \ecmpow\ governs the energy scaling of the MPI infrared regularization scale through
\begin{equation}
p_{T0}(\sqrt{s}) = p_{T0}^{\mathrm{Ref}}
\left(
\frac{\sqrt{s}}{\sqrt{s_{\mathrm{Ref}}}}
\right)^{n_{\mathrm{ecm}}}.
\label{eq:pt0_scaling}
\end{equation}
Here \(n_{\mathrm{ecm}}\) denotes the value of \ecmpow. If the same parameter point describes 7~TeV reasonably but fails to transport the soft activity to 0.9, 2.36, and 13~TeV, then \ecmpow\ is the first parameter that should be opened.

The six-parameter mixed-family scan therefore preserved the same local-basin strategy but enlarged the active parameter set minimally. The design again contained 12 points. Six were retained anchors from the best current mixed-family candidates, initialized at the Monash-like reference value \ecmpow\ = 0.215, and six were new local samples in which \ecmpow\ varied jointly with the original five-parameter core.

This extension changed the character of the fit. The grouped score dropped sharply relative to the failed five-parameter overlap stage and stabilized near
\[
S_{\mathrm{grouped}}/N_{\mathrm{bin}} \approx 273.69.
\]
A subsequent refinement restricted to the six strongest mixed-family points left the ranking unchanged. That stability matters. It showed that the gain was not a fluctuation of the wider local scan and that \ecmpow\ was not only numerically convenient. It had become physically necessary once the tune basis was promoted from a local hadronization test to a real multi-energy mixed-family objective.

\subsection{Activation of \sigmaPT}
\label{subsec:sigma_activation}

After the six-parameter mixed-family basin had stabilized, one additional hadronization degree of freedom was tested: \sigmaPT. This parameter controls the transverse width generated during string breaking and therefore provides a limited extra handle on soft hadronic broadening. Compared with \ecmpow, its expected impact is narrower, but once the multi-energy transport problem had been controlled it was reasonable to ask whether the remaining mixed-family mismatch could be reduced by a modest increase in fragmentation flexibility.

The seven-parameter scan kept the same mixed-family objective and the same local-basin philosophy. The design contained the six confirmed mixed-family anchors, each initialized at the default value \sigmaPT\ = 0.335, together with six additional local samples in which \sigmaPT\ was varied jointly with the six already active parameters. This was still not a blind global scan. It was a focused test of whether one extra soft-broadening lever improved the already viable basin.

It did, but only moderately. The best point in the seven-parameter scan became one of the newly drawn local samples rather than one of the inherited six-parameter anchors, and the best grouped score improved from about \(273.69\) to about \(273.41\). That gain was real enough to justify one further local seven-parameter refinement, but it also showed that the main mixed-family stabilization had already been achieved at six parameters. \sigmaPT\ sharpened the basin; it did not redefine it.

\subsection{Opening the MPI overlap-profile freedom}
\label{subsec:exppow_opening}

At this stage the tune had reached a stable seven-parameter mixed-family solution, and a report-level benchmark against Monash on that basis had already been established. That benchmark was useful, but it also exposed the next unresolved structure clearly: the low-energy charged-particle block remained numerically dominant, and the residual minimum-bias tension was no longer well described as a simple normalization or energy-transport problem. The natural next extension was therefore \expPow.

In the \textsc{PYTHIA} MPI model, \expPow\ controls the shape of the matter-overlap profile and therefore modifies how additional soft activity is distributed across impact parameters. This is the first parameter opened in the present study specifically to improve the structure of the minimum-bias and underlying-event sectors rather than the identified-hadron sector. Its promotion followed directly from the residual score pattern. Once \ptref\ and \ecmpow\ had already adjusted the level and energy scaling of MPI activity, the next physically sensible question was whether the overlap-profile shape itself had become the limiting factor.

The tune construction therefore continued by opening \expPow\ while preserving the existing local basin. The purpose of this step was not to restart the programme from scratch, but to test whether the unresolved low-energy and event-activity structure could be reduced by adding the first explicitly profile-sensitive MPI degree of freedom. This extension marked the point at which the tune began to move beyond the earlier mixed-family basis toward a broader soft-QCD construction.

\subsection{Adding baryon-pair production and extending the 13~TeV flavor sector}
\label{subsec:probqq_extension}

Once the MPI overlap-profile freedom had been opened, the next remaining deficiency was no longer centered on the overall level of charged-particle activity alone. The heavier identified species at 13~TeV, together with the associated flavor-composition ratios, remained an identifiable weakness. The fit basis was therefore broadened on the hadron side by adding the ALICE 13~TeV light-flavor block, and the active parameter set was enlarged by promoting \probQQtoQ.

This was a natural pairing. The new ALICE 13~TeV hadron block broadened the flavor-sensitive constraints beyond the existing CMS 13~TeV identified-hadron basis, while \probQQtoQ\ provided the first explicitly baryon-sensitive degree of freedom in the string-fragmentation sector. The purpose was not to replace the existing 13~TeV hadron measurements, but to test whether a common point could now satisfy a broader flavor basis once an explicit diquark-production lever was made available.

This step is one of the most important in the full tune construction, because it marks the point at which the Ajaz tune ceased to be only a minimum-bias plus identified-hadron compromise and became a more genuinely global soft-QCD retune. The active parameter space now contained direct controls on fragmentation shape, strangeness, baryon sensitivity, color topology, MPI normalization, MPI energy transport, MPI overlap profile, and transverse soft broadening.

\subsection{Broadening to the final global fit basis}
\label{subsec:global_basis_extension}

The final broadening of the fit basis brought the tune to its report-level global form. Four further analysis blocks were added: the CMS 7 TeV UE block, the ATLAS 7 TeV event-shape block, the ATLAS 13 TeV MB track block, and the ATLAS 13 TeV UE block.
With these additions, the fit basis no longer consisted only of charged-particle minimum-bias measurements and identified-hadron spectra. It now also contained a traditional 7~TeV underlying-event block, a 7~TeV event-shape block, a 13~TeV track-based minimum-bias block, and a 13~TeV track-based underlying-event block.

This widening changed the meaning of the tune. The earlier mixed-family benchmark had already shown that one compact seven-parameter point could improve on Monash within a narrower basis. The broader fit basis now asked a stronger question: can the same sequentially constructed tune remain competitive once topology-sensitive and track-based event-activity observables are included directly in the objective? It is this wider basis that justifies speaking of the Ajaz tune as a global soft-QCD retune within the present study.

The progression from the original five-parameter localization to the final basis matters. The final tune did not come from a single nine-dimensional scan over a broad mixed observable pool. The tune was built incrementally. At each stage, the best local basin was carried forward, new parameters were opened only when the residual score pattern made the need clear, and the fit basis was widened only after the existing basin had shown enough stability to be tested against the next sector. That is what keeps the final point readable, even though the basis ultimately becomes broad.

\subsection{Final report-level rerun and benchmark setup}
\label{subsec:report_level_setup}

Once the nine-parameter basin and the final global fit basis had stabilized, the best point was frozen and rerun at high statistics for the report-level benchmark. No new parameter exploration was performed at this stage. The purpose was confirmation, not further optimization. The same statistics policy was then applied to a Monash reference point so that the final comparison would be strictly like-for-like.

The final report-level comparison therefore used the same generator version, the same fit basis, the same merged-statistics policy, and the same grouped score definition for the Ajaz and Monash points. Each point was evaluated with
\[
24 \times 2\times10^5 = 4.8\times10^6
\]
generated events. This benchmark is the one quoted in Section~\ref{sec:results}. It is the final outcome of the full sequential construction, not an intermediate scan result.

% =========================================================
\section{The final Ajaz tune and benchmark against Monash}
\label{sec:results}

\subsection{Final tuned parameter point}
\label{subsec:final_point}

The sequential construction described in Section~\ref{sec:strategy} leads to one stable best point in the final nine-parameter subspace. This point is taken as the Ajaz tune and is benchmarked directly against Monash 2013 on the same global fit basis and on the same held-out validation basis. The tuned values are listed in Table~\ref{tab:final_tune}.

Several features of the final point are immediately visible. The longitudinal fragmentation sector remains close to Monash in \alund\ but moves more noticeably in \blund. The color-reconnection range increases strongly, the MPI regularization scale shifts upward moderately, and the energy scaling of the MPI sector becomes harder through a larger \ecmpow. The final point also prefers a slightly broader matter-overlap profile than Monash through \expPow, and it moves toward slightly larger strange-diquark production through \probQQtoQ. These shifts indicate that the Ajaz tune is not simply a local fragmentation retune. It is a correlated soft-QCD solution in which fragmentation, flavor composition, color topology, and MPI activity all move together.

\begin{table}[H]
\centering
\caption{Final tuned parameter values of the Ajaz tune compared to the Monash 2013 baseline.}
\label{tab:final_tune}
\begin{tabular}{lcc}
\toprule
Parameter & Monash 2013 & Ajaz tune \\
\midrule
\alund & 0.68 & 0.68830185 \\
\blund & 0.98 & 1.2446208 \\
\crrange & 1.80 & 3.1772336 \\
\ptref & 2.28 & 2.3388855 \\
\probStoUD & 0.217 & 0.22890479 \\
\probQQtoQ & 0.081 & 0.083583974 \\
\ecmpow & 0.215 & 0.23870553 \\
\expPow & 1.85 & 2.0171851 \\
\sigmaPT & 0.335 & 0.29964282 \\
\bottomrule
\end{tabular}
\end{table}

The present point should be read with the right scope. It is the best solution found within the final global soft-QCD basis used in this paper. It is not proposed as a universal replacement for Monash in every possible sector. Its claim is narrower and more defensible: on the common fit basis and on the held-out validation basis defined in Section~\ref{sec:data}, it gives a better overall description than Monash and does so through a physically interpretable sequence of parameter promotions.

\subsection{Total fit-basis comparison}
\label{subsec:fit_basis_benchmark}

The final fit-basis comparison is performed on exactly the same global observable basis for both tunes, using the same generator version, the same analysis chain, the same merged-statistics policy, and the same grouped score definition. On that common basis the Ajaz tune gives
\[
S_{\mathrm{grouped}}/N_{\mathrm{bin}} = 85.6721,
\]
while Monash gives
\[
S_{\mathrm{grouped}}/N_{\mathrm{bin}} = 99.1818.
\]
The improvement is therefore
\[
\Delta\!\left(S_{\mathrm{grouped}}/N_{\mathrm{bin}}\right)=99.1818-85.6721=13.5097.
\]
This is a clear separation on the common fit basis. It is much larger than the gain seen in the earlier restricted mixed-family benchmark and it survives the transition to the broader global-basis objective.

\begin{table}[H]
\centering
\caption{Total fit-basis comparison between the Ajaz tune and Monash 2013 on the common global soft-QCD basis.}
\label{tab:total_fit_benchmark}
\begin{tabular}{lcc}
\toprule
Tune & $S_{\mathrm{grouped}}/N_{\mathrm{bin}}$ & Better score \\
\midrule
Monash 2013 & 99.1818 &  \\
Ajaz tune & 85.6721 & yes \\
\bottomrule
\end{tabular}
\end{table}

Table~\ref{tab:total_fit_benchmark} reports the global weighted score density over the full fit basis. The
corresponding blockwise weighted score densities are shown separately in Table~\ref{tab:analysiswise_fit}.
The significance of this result is not only that the total score improves. It is that the improvement is distributed over several physically distinct sectors. The Ajaz tune is not winning on one isolated histogram while deteriorating everywhere else. It is stronger across most of the 13~TeV activity sector, in the ALICE identified-hadron blocks, and in several minimum-bias observables, while Monash retains an advantage only in a smaller subset of 7~TeV UE and event-shape observables.

\subsection{Analysis-wise fit-basis comparison}
\label{subsec:analysiswise_fit}

The block-by-block comparison is summarized in Table~\ref{tab:analysiswise_fit}. The entries are quoted
as blockwise weighted $S/N_{\mathrm{bin}}$ values, evaluated separately for each fitted analysis block. They
therefore show the average score density within each block rather than the raw blockwise score sum. This
distinction is important for interpreting the relation between Tables~\ref{tab:total_fit_benchmark}
and~\ref{tab:analysiswise_fit}: Table~\ref{tab:total_fit_benchmark} gives the global weighted
$S_{\mathrm{grouped}}/N_{\mathrm{bin}}$ over the full fit basis, whereas Table~\ref{tab:analysiswise_fit} resolves that
comparison block by block. The pattern is straightforward. Ajaz gives the lower score in seven of the nine fitted analyses, with the largest gains in the sectors that probe high-energy soft activity most directly.

\begin{table}[H]
\centering
\caption{Blockwise weighted ($S/N_{\mathrm{bin}}$) values in the final fit-basis benchmark. Lower values indicate better agreement. The short analysis labels are defined in Table~\ref{tab:analysis_map}.}
\label{tab:analysiswise_fit}
\begin{tabularx}{\textwidth}{>{\raggedright\arraybackslash}X cc >{\centering\arraybackslash}p{0.16\textwidth}}
\toprule
Analysis & Monash 2013 & Ajaz tune & Better tune \\
\midrule
ALICE charged-particle & 699.426 & 696.777 & Ajaz tune \\
ALICE 7 TeV identified hadrons & 14.102 & 8.558 & Ajaz tune \\
CMS 7 TeV UE & 86.400 & 90.788 & Monash 2013 \\
ATLAS 7 TeV event shapes & 18.519 & 57.250 & Monash 2013 \\
CMS 13 TeV charged density & 26.410 & 3.423 & Ajaz tune \\
CMS 13 TeV identified hadrons & 7.186 & 5.943 & Ajaz tune \\
ALICE 13 TeV light flavor & 33.479 & 25.916 & Ajaz tune \\
ATLAS 13 TeV MB tracks & 26.814 & 15.067 & Ajaz tune \\
ATLAS 13 TeV UE & 99.675 & 35.664 & Ajaz tune \\
\bottomrule
\end{tabularx}
\end{table}

Three aspects of Table~\ref{tab:analysiswise_fit} are worth noting. First, the largest single gain appears in the CMS 13~TeV charged-density block, where the score falls from 26.410 to 3.423. This is strong evidence that the final MPI sector, including \ptref, \ecmpow, and \expPow, transfers more successfully to the highest energy included in the fit. Second, the ALICE 7~TeV identified-hadron block remains one of the clearest successes of the tune, with the score decreasing from 14.102 to 8.558. Third, the broader 13~TeV additions justify their inclusion: the ALICE 13~TeV light-flavor block, the ATLAS 13~TeV minimum-bias track block, and especially the ATLAS 13~TeV UE block all improve substantially.

The weak points are equally visible. Monash remains better in the CMS 7~TeV UE block and, more strongly, in the ATLAS 7~TeV event-shape block. In addition, the ALICE charged-particle block still dominates the total score numerically for both tunes. The final separation between Ajaz and Monash is therefore limited less by the sectors where Ajaz performs well and more by one persistent low-energy charged-particle tension together with two 7~TeV topology-sensitive blocks where Monash still has the advantage.

\subsection{Held-out validation benchmark}
\label{subsec:validation_results}

The fit-basis benchmark alone is not enough for a tune paper. The tuned point must also be tested on analyses that were not used during optimization. The held-out validation basis defined in Section~\ref{subsec:validation_basis} provides that test. The result is summarized in Table~\ref{tab:validation_summary}. On the combined validation basis the Ajaz tune gives
\[
S_{\mathrm{grouped}}/N_{\mathrm{bin}} = 63.1474,
\]
while Monash gives
\[
S_{\mathrm{grouped}}/N_{\mathrm{bin}} = 68.9437.
\]
The absolute scores are still large, which is not surprising for broad soft-QCD validation blocks, but the relative trend is clear: the Ajaz tune also generalizes better than Monash to the neighboring analyses that were kept out of the fit.

\begin{table}[H]
\centering
\caption{Held-out validation benchmark. These analyses were excluded from the fit objective and are listed in Table~\ref{tab:validation_basis}.}
\label{tab:validation_summary}
\begin{tabular}{lcc}
\toprule
Tune & Total grouped score $S_{\mathrm{grouped}}$ & $S_{\mathrm{grouped}}/N_{\mathrm{bin}}$ \\
\midrule
Ajaz tune & 478909.539 & 63.1474 \\
Monash 2013 & 523489.735 & 68.9437 \\
\bottomrule
\end{tabular}
\end{table}

Table~\ref{tab:validation_breakdown} lists the grouped-score contributions of the held-out validation blocks. These entries are not normalized by the blockwise number of populated bins. They are reported to show which validation sectors drive the total separation between the Ajaz tune and Monash 2013.

The validation breakdown in Table~\ref{tab:validation_breakdown} shows that the improvement is not uniform across all held-out blocks. Monash remains slightly better in the two CMS low-energy charged-particle validation sets, consistent with the low-energy tension that also persists in the fit basis. By contrast, the Ajaz tune performs better in the ATLAS early underlying-event block and much better in the two held-out 13~TeV blocks. This pattern is important. It shows that the improvement observed in the fitted 13~TeV minimum-bias and underlying-event sectors is not limited to the fitted histograms themselves. It extends to nearby analyses that did not enter the tune selection.

\begin{table}[H]
\centering
\caption{Blockwise grouped-score contributions in the held-out validation benchmark. Lower is better.}
\label{tab:validation_breakdown}
\begin{tabular}{lcc}
\toprule
Held-out analysis & Ajaz tune & Monash 2013 \\
\midrule
CMS low-energy charged-particle block & 82171.190 & 82027.850 \\
CMS low-energy NSD block & 314062.220 & 312340.150 \\
ATLAS early UE block & 64303.317 & 72165.683 \\
ATLAS 13 TeV charged-track validation block & 17528.822 & 54190.497 \\
CMS 13 TeV charged-final-state block & 843.991 & 2765.554 \\
\bottomrule
\end{tabular}
\end{table}

The validation result strengthens the interpretation of the tune. The improvement is not confined to the chosen fit basis. It also extends to several adjacent measurements outside that basis, especially at 13~TeV. This does not make the tune universal, but it does strengthen the final benchmark.

\subsection{Representative distributions}
\label{subsec:representative_plots}

\begin{figure}[H]
    \centering
    \includegraphics[width=0.48\textwidth]{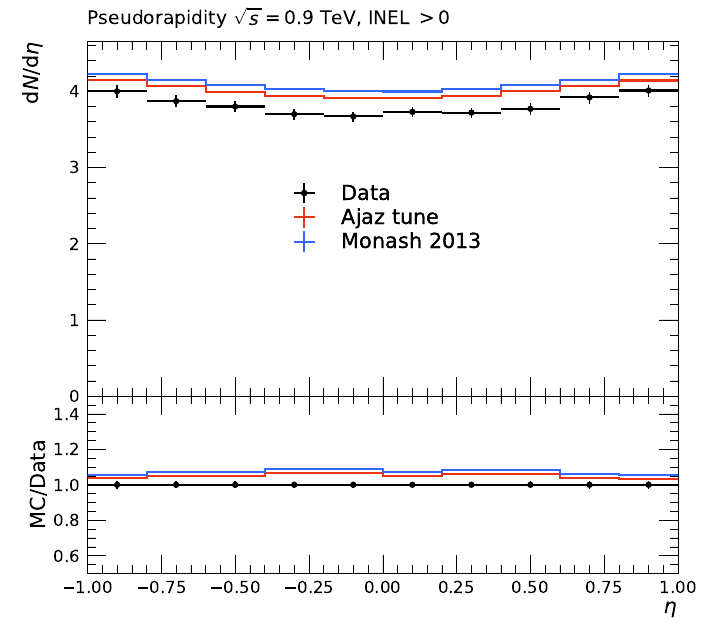}
    \includegraphics[width=0.48\textwidth]{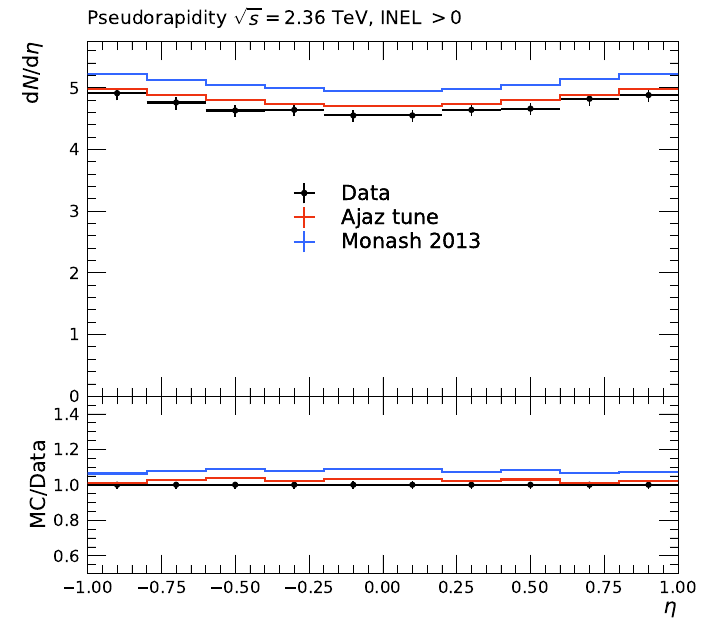}
    \includegraphics[width=0.48\textwidth]{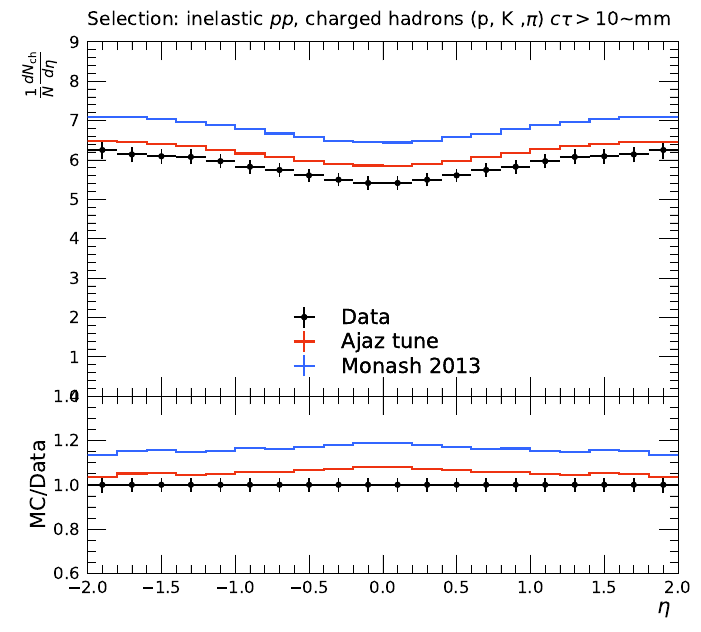}
    \includegraphics[width=0.48\textwidth]{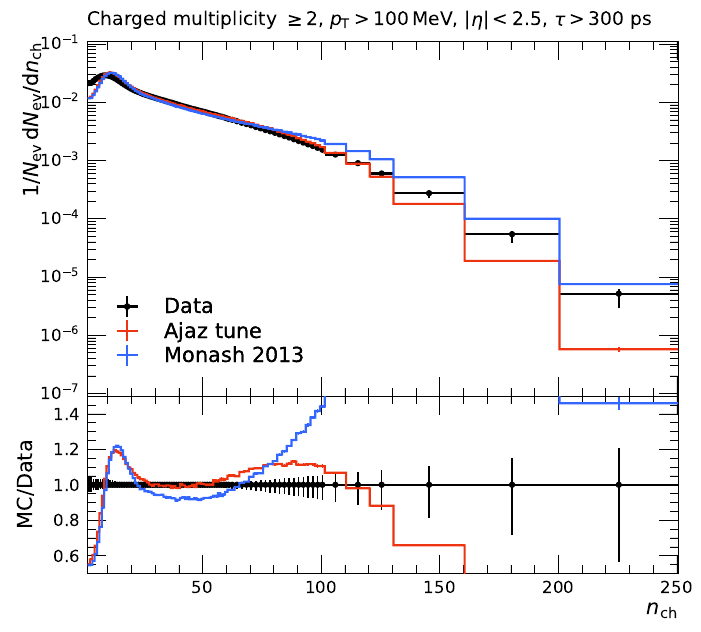}
    \caption{Representative minimum-bias observables on the final fit basis: low-energy ALICE charged-particle pseudorapidity distributions at 0.9 and 2.36~TeV, the CMS 13~TeV charged-density observable, and the ATLAS 13~TeV minimum-bias track multiplicity distribution. Ajaz improves the high-energy minimum-bias sector clearly, while the low-energy ALICE tension remains substantial for both tunes.}
    \label{fig:global_mb}
\end{figure}

\begin{figure}[H]
    \centering
    \includegraphics[width=0.48\textwidth]{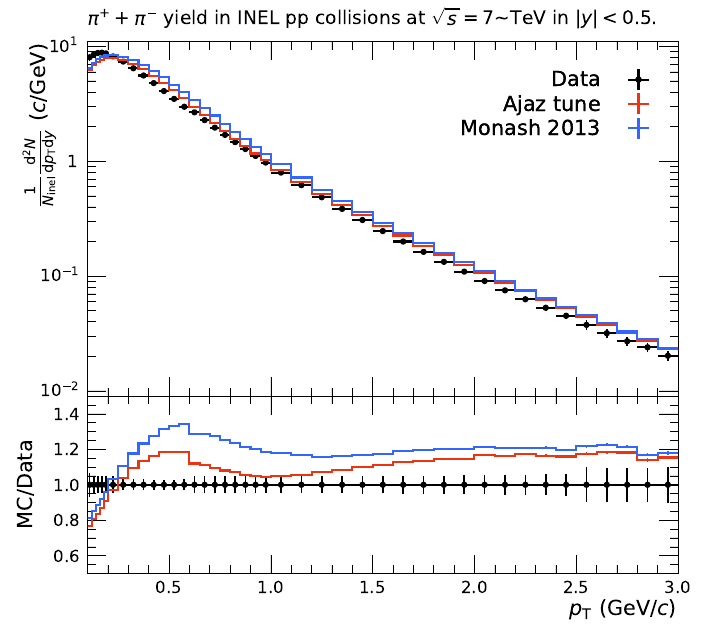}
    \includegraphics[width=0.48\textwidth]{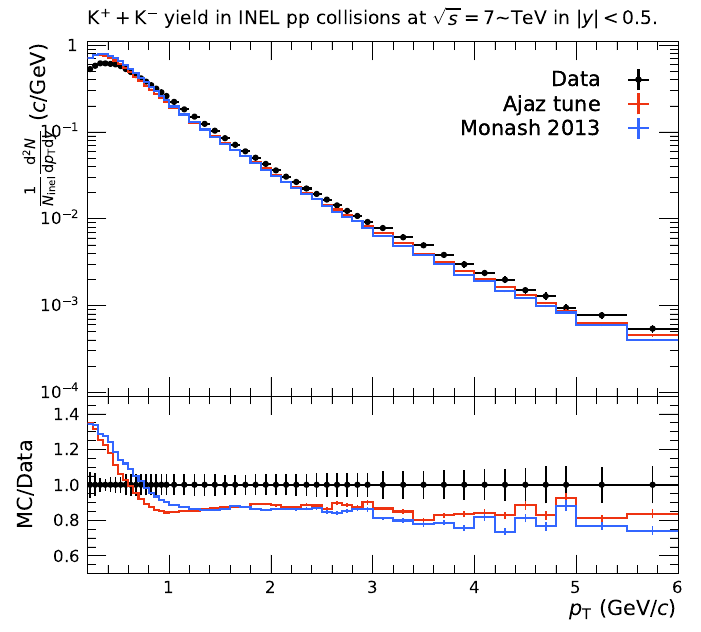}
    \includegraphics[width=0.48\textwidth]{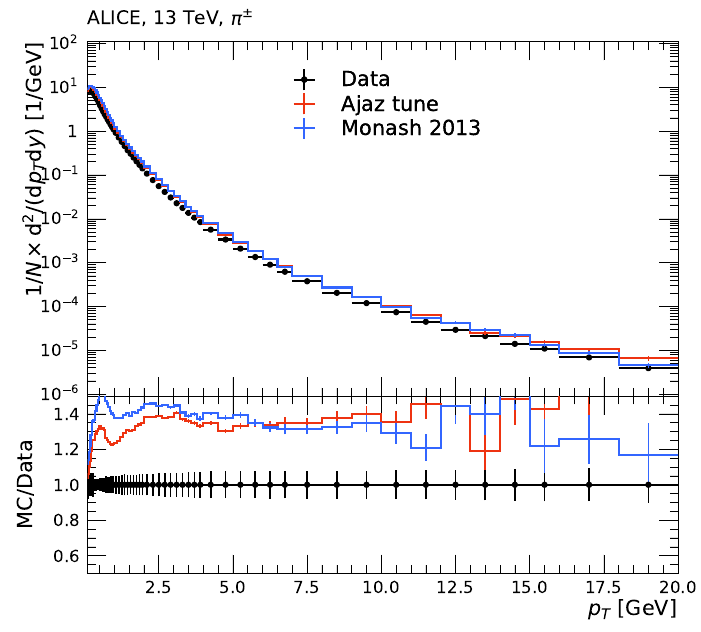}
    \includegraphics[width=0.48\textwidth]{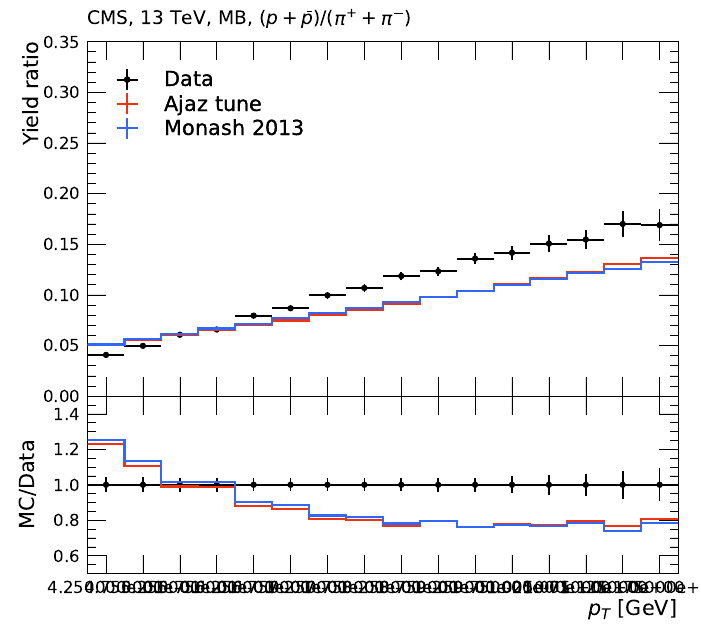}
    \caption{Representative identified-hadron observables on the final fit basis: ALICE 7~TeV pion and kaon spectra, the ALICE 13~TeV pion spectrum, and the CMS 13~TeV $(p+\bar p)/(\pi^+ + \pi^-)$ ratio. Ajaz improves the ALICE hadron sectors clearly, while Monash remains stronger in at least part of the heavier-species composition structure of the CMS 13~TeV block.}
    \label{fig:global_hadrons}
\end{figure}

\begin{figure}[H]
    \centering
    \includegraphics[width=0.48\textwidth]{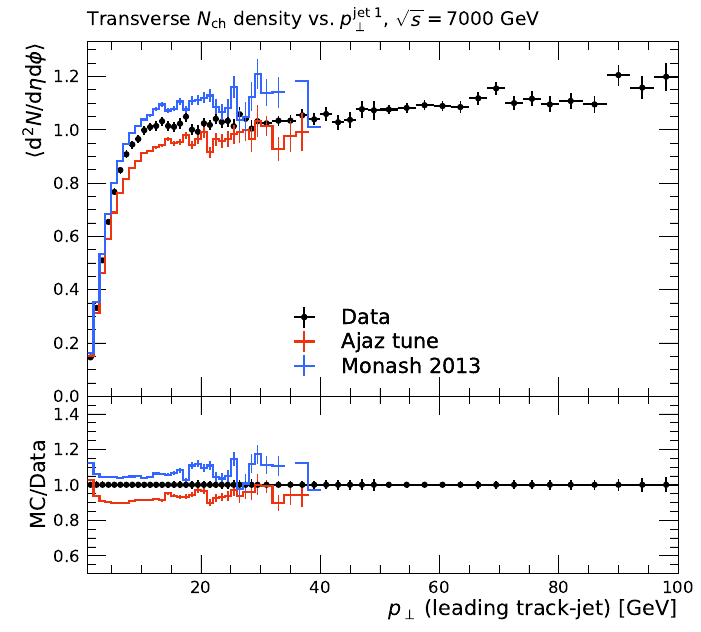}
    \includegraphics[width=0.48\textwidth]{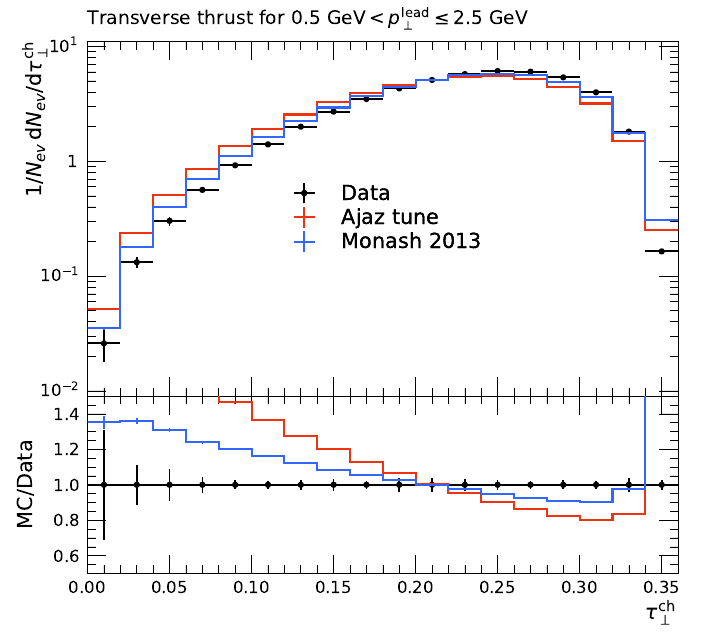}
    \includegraphics[width=0.48\textwidth]{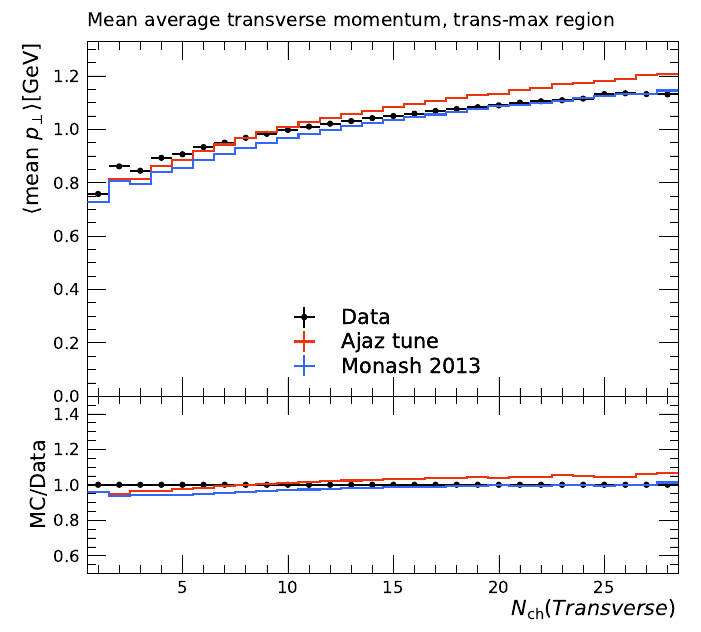}
    \includegraphics[width=0.48\textwidth]{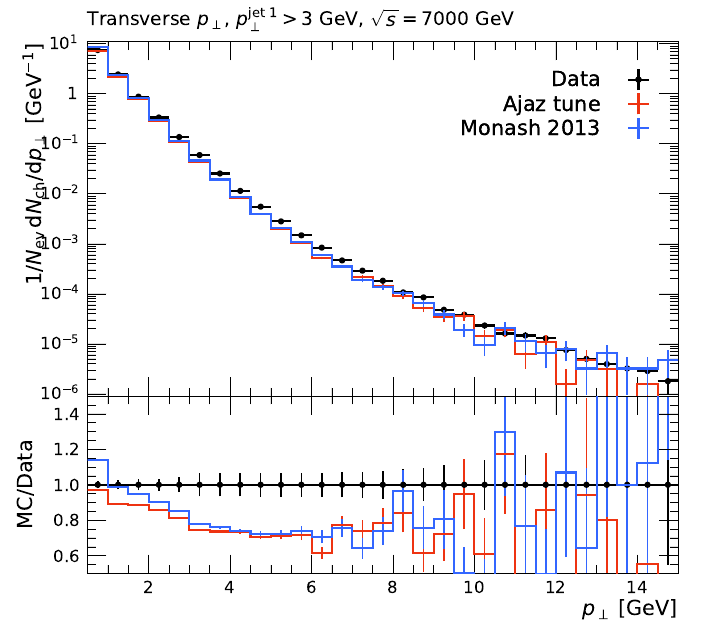}
    \caption{Representative UE and event-shape observables on the final fit basis: a CMS 7~TeV transverse charged-density distribution, an ATLAS 7~TeV transverse-thrust distribution, an ATLAS 13~TeV trans-max mean-$p_{\mathrm{T}}$ observable, and a CMS 7~TeV transverse scalar-$p_{\mathrm{T}}$ density. Ajaz improves the 13~TeV ATLAS UE sector, while Monash remains better in the CMS 7~TeV UE block and the ATLAS 7~TeV event-shape block.}
    \label{fig:global_ue_shapes}
\end{figure}

Representative distributions are shown in Figures~\ref{fig:global_mb}--\ref{fig:global_ue_shapes}. They do not replace the full block-by-block tables. They are included to show how the global benchmark looks at the histogram level.
Figure~\ref{fig:global_mb} summarizes the minimum-bias charged-particle sector. The first two panels show the low-energy ALICE pseudorapidity densities at 0.9 and 2.36~TeV, where both tunes still struggle and where the Ajaz gain remains small. The last two panels show the CMS 13~TeV charged-density observable and the ATLAS 13~TeV minimum-bias track observable. These make the main point of the high-energy activity sector clear: Ajaz transfers to the 13~TeV minimum-bias region much more successfully than Monash.

Figure~\ref{fig:global_hadrons} shows representative identified-hadron observables. The ALICE 7~TeV pion and kaon spectra demonstrate the improvement of the 7~TeV identified-hadron sector that originally anchored the early tuning stages. The ALICE 13~TeV pion spectrum shows that the tune remains competitive when the flavor basis is broadened at the same energy. The CMS 13~TeV proton-to-pion ratio illustrates the opposite side of the picture: Monash remains competitive or better in some heavier-species composition observables, even though the net CMS 13~TeV identified-hadron block is improved by Ajaz.

Figure~\ref{fig:global_ue_shapes} collects representative UE and event-shape distributions. This figure is included deliberately because the final tune does not improve every sector. The ATLAS 13~TeV UE observable is much better described by Ajaz, whereas the CMS 7~TeV UE and ATLAS 7~TeV event-shape examples still favor Monash. Those residual tensions are an essential part of the paper and help define the scope of the tune honestly.

% =========================================================
\section{Discussion}
\label{sec:discussion}

\subsection{Where the Ajaz tune improves}
\label{subsec:where_improves}

The final benchmark shows that the Ajaz tune improves the global fit basis in a patterned way rather than through many tiny accidental shifts. The strongest gains are concentrated in the high-energy minimum-bias and underlying-event sectors, with clear support from the identified-hadron blocks as well. This matters because the tune was not built as a blind nine-parameter scan. The later parameter openings were motivated by specific tensions, and the final results reflect those motivations directly.

The clearest improvements appear at 13~TeV. In the CMS 13~TeV charged-density block, the score drops from 26.410 for Monash to 3.423 for Ajaz. In the ATLAS 13~TeV minimum-bias track block, it drops from 26.814 to 15.067. In the ATLAS 13~TeV UE block, it drops from 99.675 to 35.664. These are not marginal changes. They show that the retuned MPI scale, its energy dependence, and the added overlap-profile freedom work together to improve the transfer of the soft-activity description to the highest-energy minimum-bias and UE observables in the basis.

The identified-hadron sector also gives strong support to the Ajaz point. In the ALICE 7~TeV identified-hadron block, the score decreases from 14.102 to 8.558. In the ALICE 13~TeV light-flavor block, it decreases from 33.479 to 25.916. The CMS 13~TeV identified-hadron block also improves, though more modestly, from 7.186 to 5.943. These gains show that the final point is not limited to global multiplicity or UE activity. It also remains sensitive to hadronization and flavor structure.

The validation basis reinforces this reading. Ajaz is substantially better in the ATLAS early UE block and in both 13~TeV held-out blocks. This is especially important because those analyses were not part of the fit objective. The final point is therefore doing more than fitting the exact histograms that selected it. It also improves neighboring observables, especially in the same high-energy soft-activity region where the fit-basis gains are strongest.

\subsection{Where Monash remains better}
\label{subsec:where_monash_better}

Monash remains superior in two fitted sectors, and one broader low-energy charged-particle tension still limits both tunes.

Inside the fit basis, Monash remains better in the CMS 7~TeV UE block and in the ATLAS 7~TeV event-shape block. The event-shape difference is not small. This suggests that the final Ajaz point, while clearly stronger in the high-energy activity sector, does not yet capture every aspect of the 7~TeV global event geometry equally well. The same conclusion, though more mildly, applies to the CMS 7~TeV UE structure.

The larger unresolved issue is the low-energy charged-particle sector. The ALICE charged-particle block still dominates the total score numerically for both tunes, and the Ajaz improvement there is small compared with the absolute size of the block:
\[
699.426 \rightarrow 696.777.
\]
This means that the largest remaining tension is not primarily a matter of choosing between Ajaz and Monash. It is a deeper low-energy minimum-bias problem that neither tune resolves fully within the present parameter space.

The validation basis shows the same pattern. Monash remains slightly better in the two held-out CMS low-energy charged-particle blocks. This is fully consistent with the fit-basis result and strengthens the conclusion that the low-energy part of the problem is structurally different from the high-energy sectors where Ajaz improves clearly.

\subsection{Physical reading of the parameter shifts}
\label{subsec:physics_reading}

The final parameter shifts should be read qualitatively rather than one by one in isolation, because the active parameters remain correlated inside the final basis. The moderate movement in \alund\ and the larger increase in \blund\ reflect a retuning of the longitudinal fragmentation shape without a radical departure from the Monash hadronization baseline. The increase in \crrange\ points to a stronger role for reconnection in the final global solution, which is natural once identified-hadron spectra, flavor ratios, and activity observables are all constrained at the same time.

The MPI sector shifts are equally informative. The upward movement in \ptref\, together with a larger \ecmpow, indicates that the final point prefers a somewhat different normalization and energy transport of additional soft activity than Monash. The opening of \expPow\ then adds the missing profile flexibility needed to improve the 13~TeV minimum-bias and UE sectors. In other words, the high-energy gains are not coming from hadronization alone. They are tied directly to the broadening of the MPI description.

The final addition of \probQQtoQ\ has a narrower but still important role. It gives the tune a little more freedom in the heavier-species sector once the 13~TeV flavor basis is widened. The final value does not move far from Monash, which is itself informative: the resulted tune does not need a dramatic baryon-pair shift to obtain the final improvement. The gain comes mainly from the combined action of moderate changes across several sectors rather than from one extreme parameter displacement.

\subsection{Scope of the claim}
\label{subsec:scope_claim}

The scope of the present claim is intentionally narrow. The Ajaz tune is a global-basis improvement over Monash within the observable system used in this paper. It is constrained by minimum-bias charged-particle observables, identified-hadron spectra, UE measurements, event shapes, and 13~TeV track-based activity observables across four pp collision energies. On that common basis it performs better overall than Monash, and it also performs better on the held-out validation basis.

At the same time, no universality claim is made beyond the tested basis. The persistent low-energy charged-particle tension remains. Monash still performs better in the CMS 7~TeV UE block and the ATLAS 7~TeV event-shape block. The Ajaz tune should therefore be presented as a controlled global soft-QCD tune on the present basis, not as a final answer for every soft-QCD observable that one might test in \textsc{PYTHIA}.

A full tune-uncertainty envelope, alternative weighting schemes, and covariance-aware scoring were outside the scope of the present study.

% =========================================================
\section{Conclusions}
\label{sec:conclusion}

We have presented the Ajaz tune, a sequentially constructed nine-parameter retune of \textsc{PYTHIA}~8.316 constrained by a broadened soft-QCD basis in pp collisions at \(\sqrt{s}=0.9,\ 2.36,\ 7,\) and \(13\)~TeV. The tuning path was built step by step. A compact five-parameter hadronization-sensitive basin was localized first, then widened through the MPI energy-scaling and transverse-smearing parameters, and finally extended to the overlap-profile and baryon-pair sectors as the fit basis itself broadened to include underlying-event, event-shape, and 13~TeV track-based activity observables.

On the final common fit basis, the Ajaz tune improves the grouped score density relative to Monash from
\[
99.18 \rightarrow 85.67.
\]
This is a substantial gain. It is driven mainly by a much better description of the 13~TeV charged-density, minimum-bias track, and UE sectors, together with clear improvements in the ALICE 7~TeV identified-hadron block and the ALICE 13~TeV light-flavor block. The tune also performs better on the held-out validation basis, where the grouped score density improves from 68.94 for Monash to 63.15 for Ajaz. The strongest validation gains again appear in the 13~TeV and UE-sensitive observables.

The remaining weaknesses are also clear. The low-energy charged-particle sector remains difficult for both tunes and still dominates the total score numerically. Monash remains better in the CMS 7~TeV UE block and in the ATLAS 7~TeV event-shape block. Ajaz is therefore not presented here as a universal replacement for Monash in every soft-QCD application. It is presented as a broader and better-performing global soft-QCD tune on the basis tested in this paper.

The study shows that a sequential direct-generation workflow, guided by sector-level physics tensions rather than by an undifferentiated high-dimensional scan, improves the Monash baseline on a broad common soft-QCD basis while keeping the remaining tensions explicit.

% =========================================================
\appendix

\section{Supplementary scan history}
\label{app:scan_history}
This appendix summarizes the scan history used to build the tune. The main text explains why each parameter was promoted and why the basis was broadened when it was. Here the emphasis is on documenting the successive scan levels, the retained candidate sets, and the final report-level rerun.
\subsection{Stage summary}
\label{app:stage_summary}
Summary of the sequential tuning stages that led to the final Ajaz tune is given in Table \ref{tab:appendix_stage_summary}. In the table the event counts refer to the merged statistics per point.

\begin{table}[p]
\centering
\caption{Summary of the sequential tuning stages that led to the final Ajaz tune. Event counts refer to the merged statistics per point.}
\label{tab:appendix_stage_summary}
\small
\begin{tabularx}{\textwidth}{>{\raggedright\arraybackslash}p{0.18\textwidth} >{\raggedright\arraybackslash}p{0.24\textwidth} >{\centering\arraybackslash}p{0.11\textwidth} >{\centering\arraybackslash}p{0.17\textwidth} X}
\toprule
Step & Active parameters & Points & Events per point & Role \\
\midrule
Initial coarse scan & \alund, \blund, \crrange, \ptref, \probStoUD & 21 & $3.0\times10^6$ & Locate the first viable hadronization-sensitive basin \\
Local five-parameter refinement & same five parameters & 10 & $2.4\times10^6$ & Resolve the leading five-parameter basin more finely \\
Retained 7 TeV comparison & same five parameters & 6 finalists & $2.4\times10^6$ & Test the best 7 TeV basin against combined spectra and charged-particle constraints \\
Mixed-family overlap & same five parameters & 12 & stage-level & Check whether the 7 TeV basin survives the multi-energy mixed-family objective \\
Six-parameter extension & + \ecmpow & 12, then local retainers & stage-level & Open MPI energy scaling for the multi-energy basis \\
Seven-parameter extension & + \sigmaPT & 12, then local refinement & stage-level & Add limited transverse broadening in string breaking \\
Global-basis opening I & + \expPow & local retained basis & stage-level & Open overlap-profile freedom for UE and 13 TeV activity tensions \\
Global-basis opening II & + \probQQtoQ & local retained basis & stage-level & Add baryon-pair freedom while broadening the 13 TeV flavor basis \\
Final global-basis rerun & frozen Ajaz and Monash points & 2 comparison points & $4.8\times10^6$ & Final fit-basis and validation-basis benchmark \\
\bottomrule
\end{tabularx}
\end{table}

\subsection{Early five-parameter history}
\label{app:pilot_local}

The first explicit scan explored the five-dimensional subspace \{\alund,\ \blund,\ \crrange,\ \ptref,\ \probStoUD\}. Its purpose was to determine whether a compact hadronization-sensitive basis could already locate a stable candidate region in the identified-hadron sector. The initial scan contained 21 points, each evaluated with 15 replicas of $2\times10^5$ events, corresponding to $3.0\times10^6$ merged events per point. The best region was then refined locally with 10 points and 12 replicas of $2\times10^5$ events, corresponding to $2.4\times10^6$ merged events per point.

The retained local refinement box was
\begin{equation}
\begin{aligned}
\text{\texttt{StringZ:aLund}} &\in [0.48878204,\;1.2], \\
\text{\texttt{StringZ:bLund}} &\in [0.97046456,\;1.8772126], \\
\text{\texttt{ColourReconnection:range}} &\in [2.9537838,\;3.2772971], \\
\text{\texttt{MultipartonInteractions:pT0Ref}} &\in [2.313851,\;2.3910248], \\
\text{\texttt{StringFlav:probStoUD}} &\in [0.19915096,\;0.24529151].
\end{aligned}
\label{eq:app_refine_box}
\end{equation}
This box served only as a local refinement region and should not be confused with the wider prior limits used at the exploratory stage.

\subsection{From the mixed-family basis to the final global basis}
\label{app:mixed_to_global}

After the retained 7~TeV comparison, the tune was widened gradually rather than all at once. The first mixed-family extension showed that the retained five-parameter basin did not remain adequate once the low-energy charged-particle anchor and the 13~TeV charged-density block were imposed together. That motivated the opening of \ecmpow. Once the resulting six-parameter basin stabilized, \sigmaPT\ was added to test whether limited extra broadening in string breaking could improve the joint spectra-and-density description. The gain at that stage was real but modest.

The later global-basis openings were driven by the remaining pattern of deficits rather than by a desire to maximize parameter count. The overlap-profile parameter \expPow\ was opened when the broadened 13~TeV activity sector exposed structure that could not be repaired cleanly within the earlier MPI subspace. The baryon-pair parameter \probQQtoQ\ was then added when the 13~TeV flavor sector was extended further. These two openings completed the final nine-parameter Ajaz basis used for the report-level rerun.

\subsection{Report-level benchmark setup}
\label{app:report_benchmark}

The final benchmark froze the best Ajaz point and the Monash reference and reran both on the same fit basis with $24\times2\times10^5 = 4.8\times10^6$ events per tune point. The held-out validation basis was then evaluated without any further retuning. The resulting summary numbers are repeated here for convenience:
\[
S_{\mathrm{grouped}}/N_{\mathrm{bin}} = 85.6721 \quad \text{(Ajaz fit basis)},
\]
\[
S_{\mathrm{grouped}}/N_{\mathrm{bin}} = 99.1818 \quad \text{(Monash fit basis)},
\]
\[
S_{\mathrm{grouped}}/N_{\mathrm{bin}} = 63.1474 \quad \text{(Ajaz validation basis)},
\]
\[
S_{\mathrm{grouped}}/N_{\mathrm{bin}} = 68.9437 \quad \text{(Monash validation basis)}.
\]

\section{Final tune card and reproducibility details}
\label{app:cards}

\subsection{Incremental tune card}
\label{app:final_card}

The Ajaz tune is defined relative to the Monash 2013 baseline. The incremental card is therefore
\begin{verbatim}
StringZ:aLund = 0.68830185
StringZ:bLund = 1.2446208
ColourReconnection:range = 3.1772336
MultipartonInteractions:pT0Ref = 2.3388855
StringFlav:probStoUD = 0.22890479
StringFlav:probQQtoQ = 0.083583974
MultipartonInteractions:ecmPow = 0.23870553
MultipartonInteractions:expPow = 2.0171851
StringPT:sigma = 0.29964282
\end{verbatim}

\subsection{Software and analysis environment}
\label{app:software_env}

The scans and final reruns reported in this paper used \textsc{PYTHIA}~8.316 for event generation and \textsc{Rivet}~4.1.1 for generator-data comparison. All candidate points and the Monash baseline were evaluated through the same analysis chain. The fit basis and validation basis are listed in Tables~\ref{tab:fit_basis} and \ref{tab:validation_basis}, with the exact Rivet mappings given in Table~\ref{tab:analysis_map}.

\subsection{Statistics and replica policy}
\label{app:replica_policy}

The workflow used deterministic independent replicas at fixed parameter point and fixed collision energy. Replicas were merged before scoring. The main completed stages used the following policy:
\begin{itemize}
  \item initial coarse scan: 15 replicas of $2\times10^5$ events per point,
  \item local five-parameter refinement: 12 replicas of $2\times10^5$ events per point,
  \item retained 7~TeV comparison: 12 replicas of $2\times10^5$ events per point,
  \item final report-level rerun: 24 replicas of $2\times10^5$ events per point.
\end{itemize}
Exact reruns therefore require not only the incremental tune card but also the same software stack, analysis environment, and seed policy.

\subsection{Score definition and reproducibility scope}
\label{app:repro_scope}

The grouped score is built from the bin-wise quantity
\begin{equation}
\chi_i^2 =
\frac{\left(y_i^{\mc}-y_i^{\data}\right)^2}
{\left(\sigma_i^{\data}\right)^2+\left(\sigma_i^{\mc}\right)^2},
\end{equation}
summed first inside each analysis block and then across the grouped fit basis with unit analysis weights. The validation benchmark uses the same bin-wise definition on the held-out validation basis. The incremental tune card is enough to regenerate the physics point, but reproducing the exact quoted scores also requires the same statistics policy and the same analysis basis.

\section{Supplementary comparison and validation plots}
\label{app:plots}

This appendix collects supplementary Ajaz--Monash comparisons that are not shown in the main text. Their purpose is to make the score pattern discussed in Sections~\ref{sec:results} and \ref{sec:discussion} visible at the histogram level. The figures are grouped by same observable family for comparison. The Ajaz tune improves the global description through a broader gain across several connected soft-QCD sectors, while the remaining deficits stay confined to a smaller set of observables.

The appendix is divided into two parts. The first presents supplementary comparisons from the fitted global basis itself. The second presents representative panels from the held-out validation basis. Taken together, these figures show where the Ajaz tune improves on Monash 2013, where the two descriptions remain broadly comparable, and where Monash still performs better.

\subsection{Supplementary fit-basis comparisons}
\label{app:fit_plots}

The fitted global basis spans low-energy charged-particle observables, identified-hadron spectra, minimum-bias track observables, underlying-event measurements, and event shapes. The representative panels collected here are chosen to reflect that structure. They show that the gain of the Ajaz tune is not confined to one isolated observable family, but instead arises from a broader pattern across several soft-QCD sectors, especially the 13~TeV minimum-bias and underlying-event observables and the identified-hadron sector.

\subsubsection{Low-energy charged-particle sector}

\begin{figure}[H]
  \centering
  \includegraphics[width=0.48\textwidth]{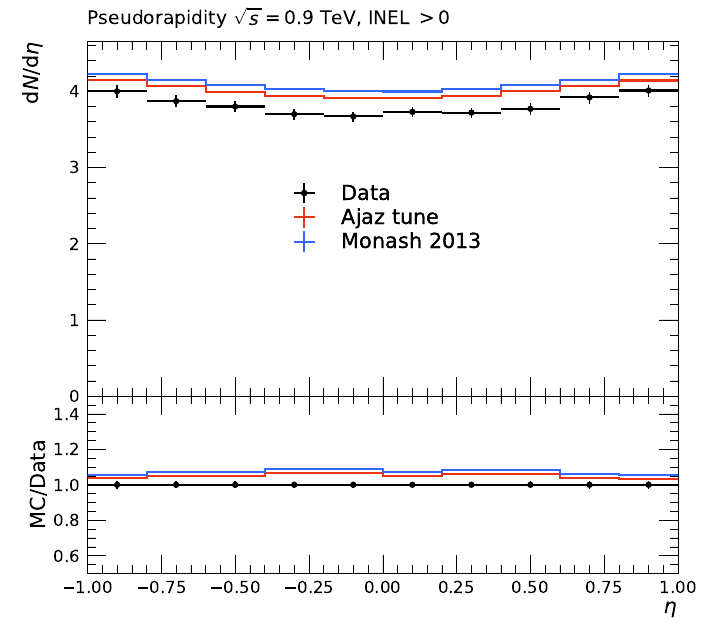}
  \includegraphics[width=0.48\textwidth]{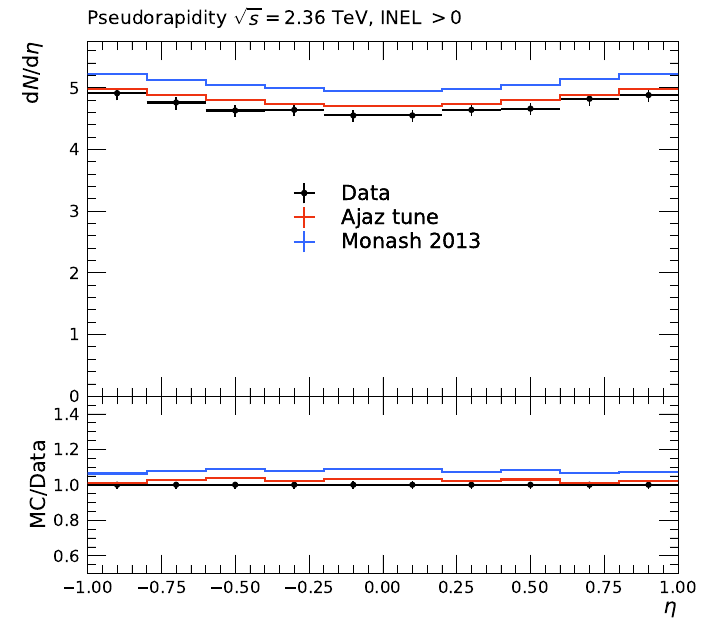}
  \caption{Representative low-energy charged-particle distributions from the ALICE charged-particle block at \(\sqrt{s}=0.9\) and \(2.36\)~TeV. These panels illustrate the sector that continues to dominate the total score for both tunes.}
  \label{fig:app_low_energy_mb}
\end{figure}

The low-energy charged-particle pseudorapidity and multiplicity distributions remain the most difficult part of the fit for both tunes. Representative panels are shown in Fig.~\ref{fig:app_low_energy_mb}. These distributions matter because they continue to dominate the remaining global tension even after the tuning was extended beyond the initial mixed-family basis.

The visual pattern in this sector should be read conservatively. The Ajaz tune does not remove the main low-energy minimum-bias tension. What it does achieve is a broadly comparable description in this sector while preserving the stronger gains seen elsewhere in the global fit basis. The low-energy charged-particle block therefore remains the main common deficit of both tunes rather than the sector that determines their separation.

\subsubsection{Identified-hadron sector}

\begin{figure}[H]
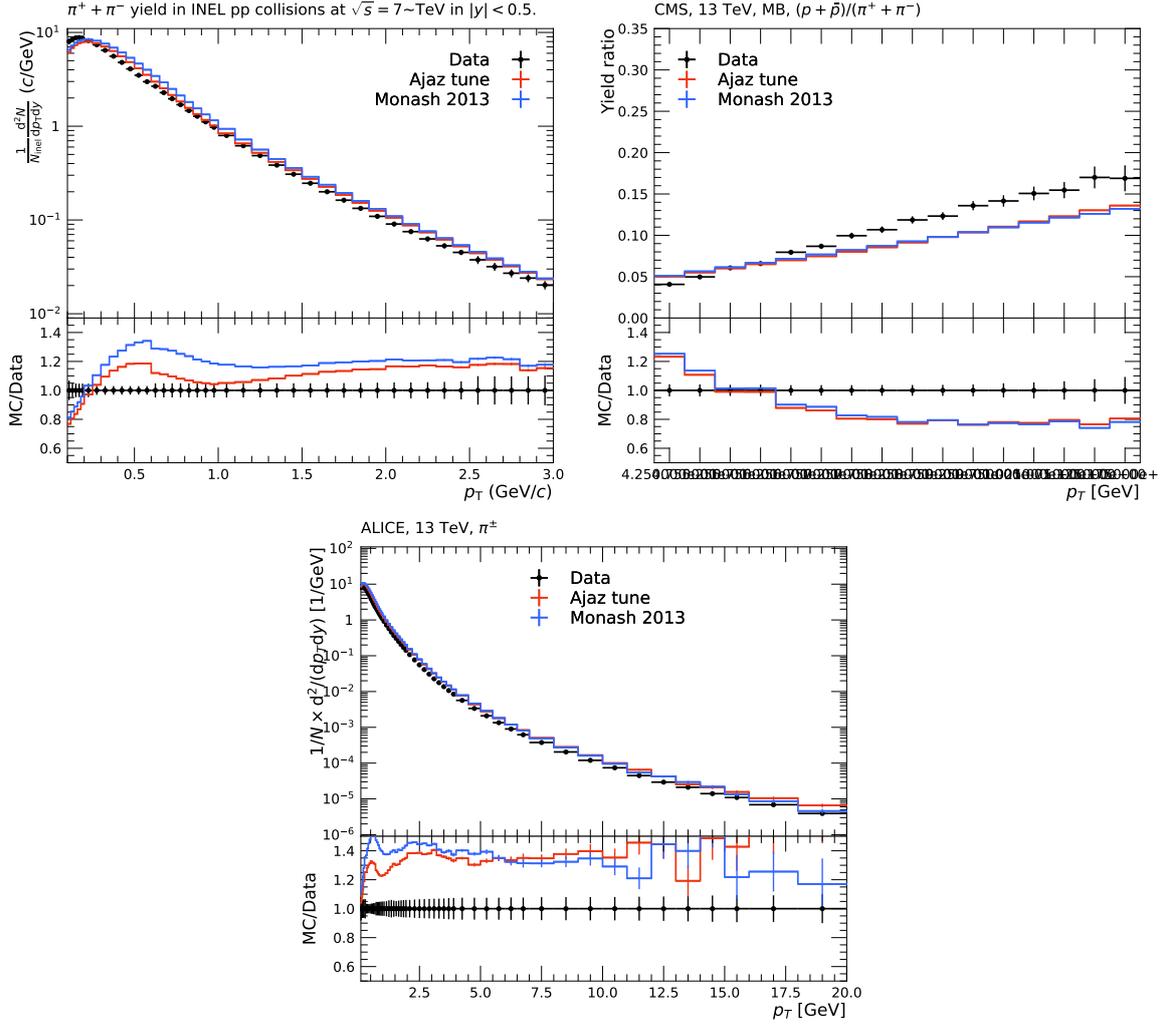

  \centering
  \includegraphics[width=0.48\textwidth]{appendix_all/fig03_E7000_ALICE_2015_I1357424_d01-x01-y01.pdf}
  \includegraphics[width=0.48\textwidth]{appendix_all/fig07_E13000_CMS_2017_I1608166_d47-x01-y01.pdf}
  \includegraphics[width=0.48\textwidth]{appendix_all/fig08_E13000_ALICE_2020_I1797443_d01-x01-y01.pdf}
  \caption{Representative identified-hadron comparisons at 7 and 13~TeV. The grouped panels show the 7~TeV ALICE hadron-spectrum sector together with 13~TeV CMS and ALICE hadron-sector distributions, illustrating the broader flavor-sector improvement of the Ajaz tune at high energy.}
  \label{fig:app_identified_hadrons}
\end{figure}

The identified-hadron measurements remain one of the clearest strengths of the Ajaz tune. Figure~\ref{fig:app_identified_hadrons} groups representative 7~TeV and 13~TeV hadron-sector panels from the ALICE and CMS measurements included in the fit basis. These comparisons show that the tuned fragmentation, color-reconnection, and flavor-sensitive parameters respond coherently once constrained on the broader global basis.

The supplementary panels also help clarify the balance of the final hadron-sector result. The improvement is broad in the ALICE hadron blocks and remains visible in part of the 13~TeV flavor-sensitive structure. At the same time, the CMS 13~TeV hadron sector is more mixed, with Monash still competitive in part of the heavier-species composition pattern. The role of these figures is therefore not to suggest uniform superiority in all hadron observables, but to show the wider and more balanced improvement across the identified-hadron basis.

\subsubsection{Minimum-bias, underlying-event, and event-shape sectors}

\begin{figure}[H]
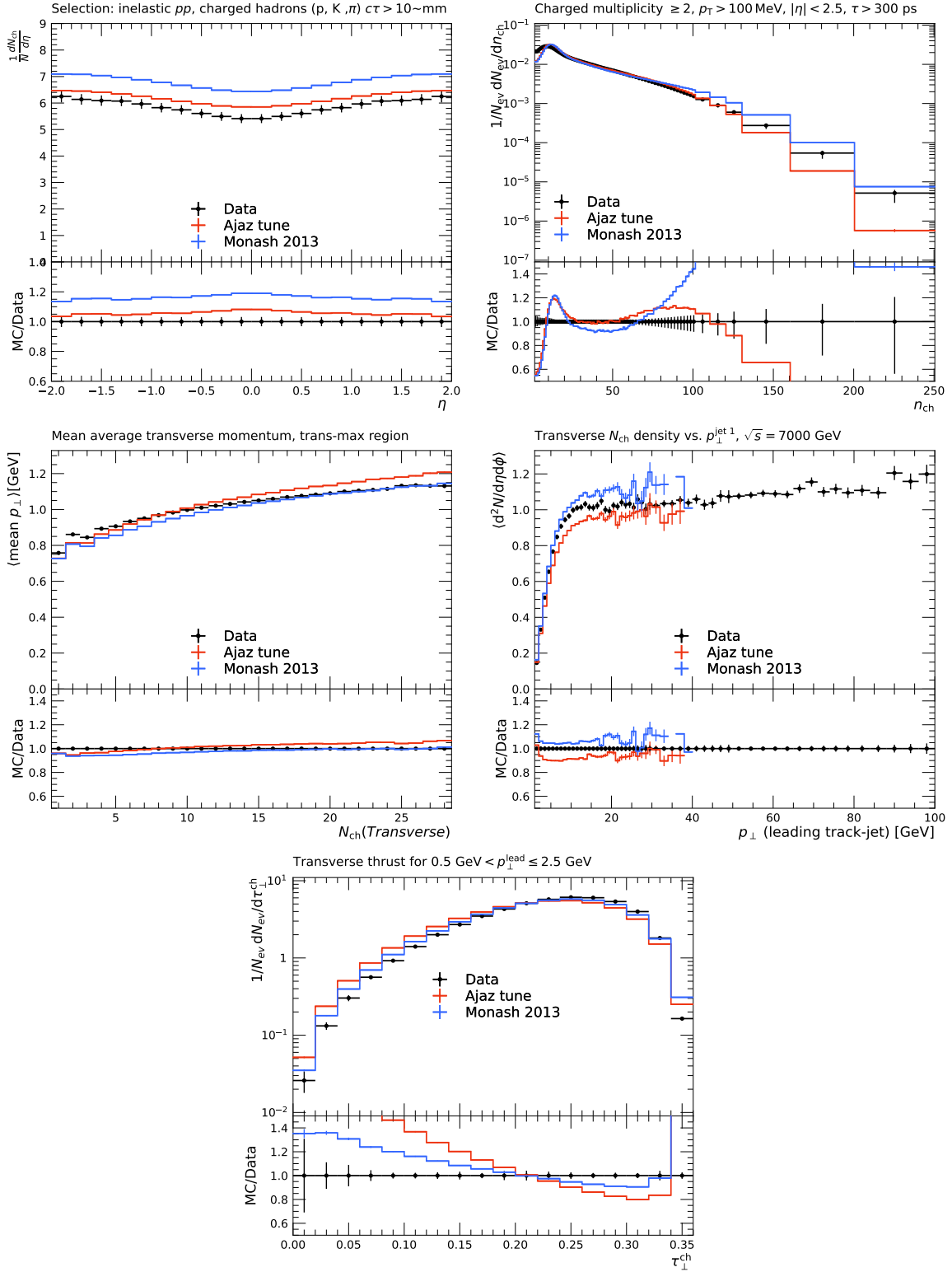

  \centering
  \includegraphics[width=0.48\textwidth]{appendix_all/fig06_E13000_CMS_2015_I1384119_d01-x01-y01.pdf}
  \includegraphics[width=0.48\textwidth]{appendix_all/fig09_E13000_ATLAS_2016_I1467230_d01-x01-y01.pdf}
  \includegraphics[width=0.48\textwidth]{appendix_all/fig10_E13000_ATLAS_2017_I1509919_d10-x01-y01.pdf}
  \includegraphics[width=0.48\textwidth]{appendix_all/fig04_E7000_CMS_2011_I916908_d01-x01-y01.pdf}
  \includegraphics[width=0.48\textwidth]{appendix_all/fig05_E7000_ATLAS_2012_I1124167_d01-x01-y01.pdf}
  \caption{Grouped minimum-bias, underlying-event, and event-shape comparisons. The upper row shows representative 13~TeV CMS and ATLAS charged-particle and underlying-event sectors in which the Ajaz tune is clearly stronger, while the lower row shows the 7~TeV CMS underlying-event and ATLAS event-shape sectors in which Monash still remains favored.}
  \label{fig:app_mb_ue_eventshape}
\end{figure}

The broader gain of the Ajaz tune is most visible in the 13~TeV minimum-bias and underlying-event sectors. Figure~\ref{fig:app_mb_ue_eventshape} groups representative 13~TeV charged-particle density, minimum-bias, and underlying-event panels together with the principal 7~TeV residual tensions. This layout is useful because it shows, in one place, both the main sources of improvement and the main sectors in which Monash remains stronger.

The 13~TeV panels correspond to observables for which the Ajaz tune gives visibly better agreement than Monash and that contribute significantly to the improved global score. By contrast, the 7~TeV panels represent the underlying-event and event-shape sectors in which Monash remains favored. These residual tensions matter because they show that the Ajaz tune is not uniformly better in every topology-sensitive observable. Its advantage is instead the result of a broader net improvement across the global fit basis.

\subsection{Supplementary held-out validation comparisons}
\label{app:validation_plots}

The held-out validation benchmark provides an external check on whether the improvement of the Ajaz tune over Monash transfers beyond the fitted global basis. These distributions were excluded from the fit objective itself and therefore test the tuned point in a more independent way.

The validation basis contains the CMS low-energy charged-particle blocks, the ATLAS early underlying-event block, the ATLAS 13~TeV charged-track validation block, and the CMS 13~TeV charged-final-state block. On the combined held-out benchmark, the Ajaz tune again performs better than Monash,
\[
S_{\mathrm{grouped}}/N_{\mathrm{bin}}:
\qquad
68.9437 \;\rightarrow\; 63.1474.
\]
The panels shown below are chosen to expose the structure of this result rather than to duplicate every available validation distribution.

\subsubsection{Validation panels with mixed behavior}

\begin{figure}[H]
  \centering
  \includegraphics[width=0.48\textwidth]{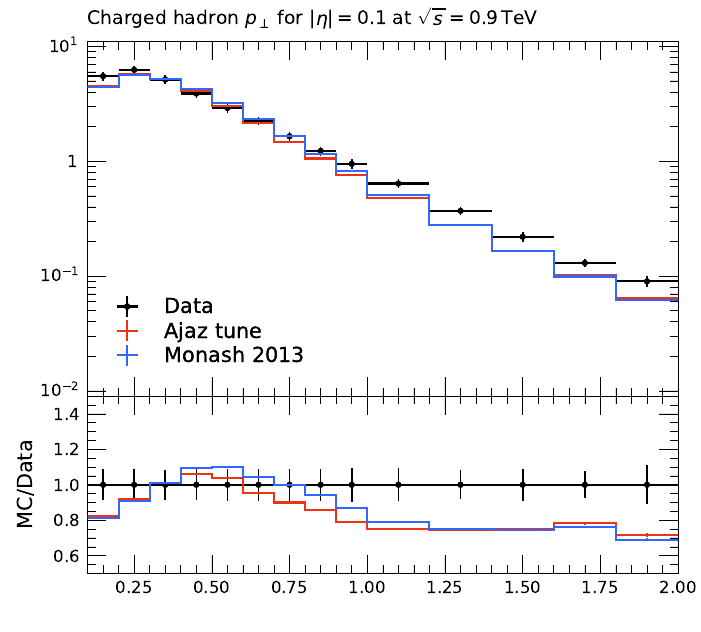}
  \includegraphics[width=0.48\textwidth]{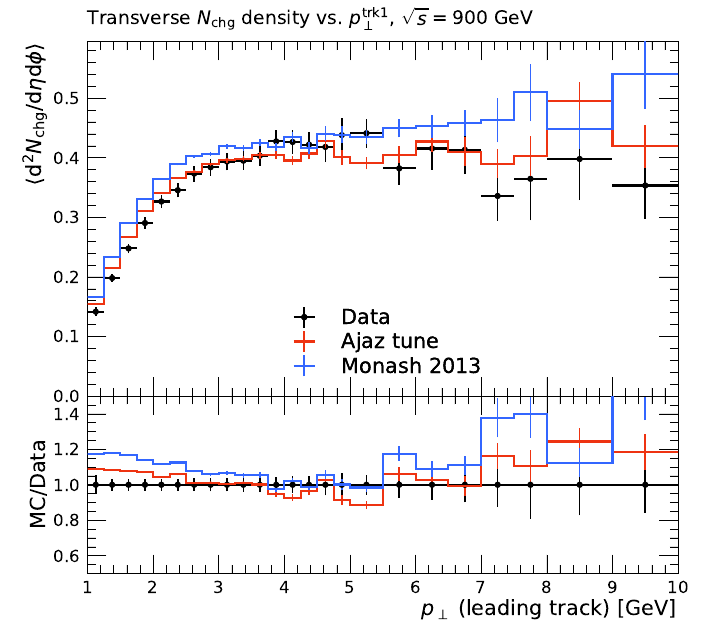}
  \includegraphics[width=0.48\textwidth]{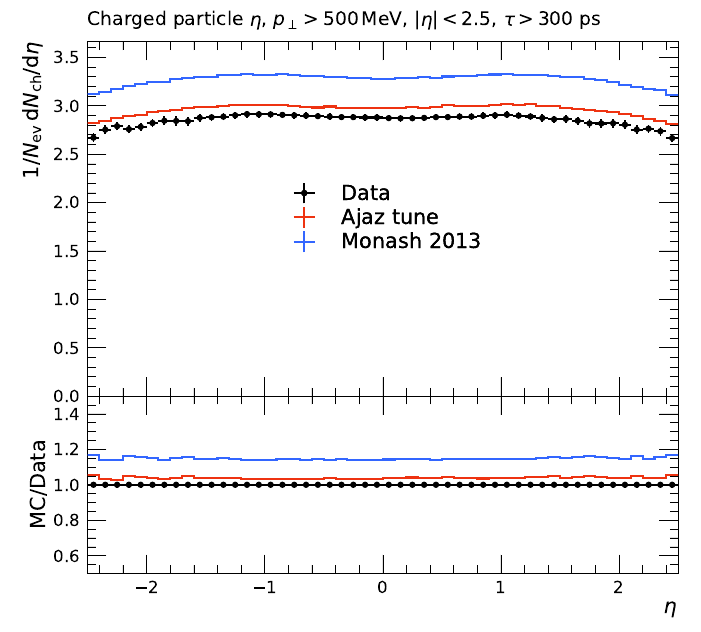}
  \includegraphics[width=0.48\textwidth]{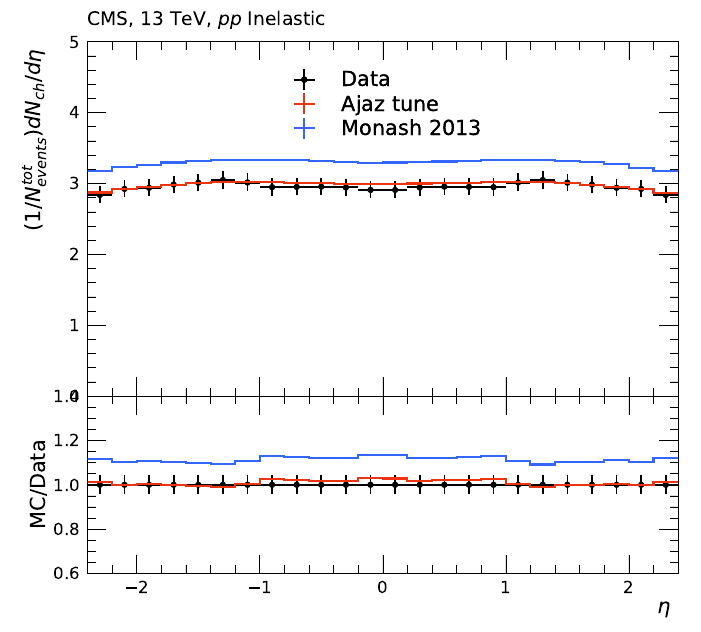}
  \includegraphics[width=0.48\textwidth]{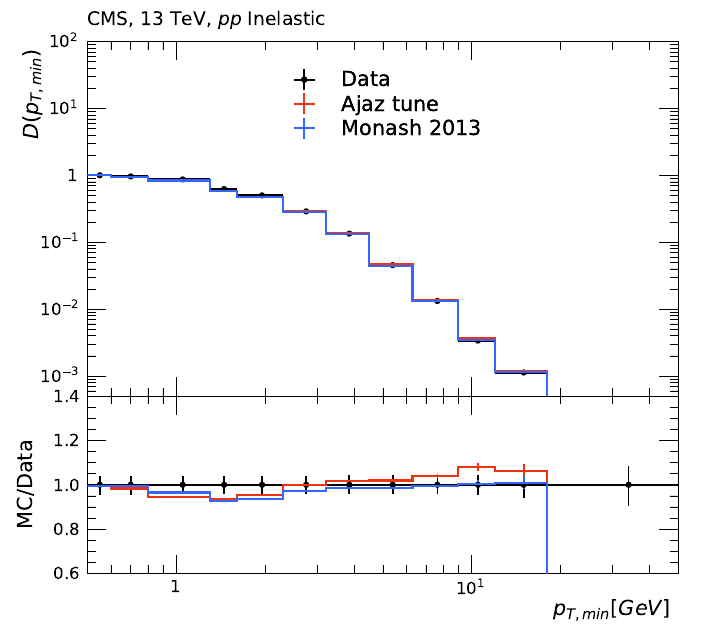}
  \caption{Grouped held-out validation comparisons across 0.9, 7, and 13~TeV. The upper row shows representative CMS minimum-bias panels in which Monash remains slightly favored, while the lower rows show ATLAS and CMS validation sectors in which the Ajaz tune is clearly stronger. Together these panels explain why the held-out validation benchmark still favors the Ajaz tune overall.}
  \label{fig:app_validation_all}
\end{figure}

The validation pattern is not uniform across all held-out sectors. In part of the CMS low-energy minimum-bias validation basis, Monash remains slightly favored. By contrast, the strongest held-out gains appear in the ATLAS and high-energy CMS validation sectors. Figure~\ref{fig:app_validation_all} groups representative panels from both categories so that the origin of the total validation separation can be seen directly.

This figure supports the same interpretation as the validation table in the main text. Monash remains slightly better in the low-energy CMS validation blocks, consistent with the same low-energy tension already visible in the fit basis. The Ajaz tune, however, performs better in the ATLAS early underlying-event block and much better in the held-out 13~TeV validation sectors. The validation gain is therefore not accidental and is not confined to the fitted histograms themselves. It extends to nearby analyses that were not used in the tune selection.

The supplementary figures support the same conclusion as the main text. The Ajaz tune does not gain its advantage from one isolated corner of the observable basis. Its improvement is broader and physically coherent, with the strongest gains in the identified-hadron, 13~TeV minimum-bias, and 13~TeV underlying-event sectors, while the remaining tensions stay localized in the low-energy charged-particle block and a smaller number of 7~TeV topology-sensitive observables.

\section*{Acknowledgment:} The authors extend their appreciation to the Deanship of Scientific Research and Libraries in Princess Nourah bint Abdulrahman University for funding this research work through the Supporting Publication in Top-Impact Journals Initiative (SPTIF-2026).
\section*{Author Contributions:} All authors listed have made a substantial, direct, and intellectual contribution to the work and approved it for publication.
\section*{Data Availability Statement:} the experimental data used to support the findings of this study are included within the article and are cited at relevant places within the text as references.
\section*{Compliance with Ethical Standards:} The authors declare that they are in compliance with ethical standards regarding the content of this paper.
\section*{Conflict of Interest:} The authors declare that there are no conflicts of interest regarding the publication of this paper.
\\
\bibliographystyle{elsarticle-num}
\bibliography{references}

\end{document}